\documentclass[10pt, pre, amsmath, onecolumn, showpacs, superscriptaddress]{revtex4-1}
\usepackage{graphicx,color}
 \graphicspath{{./}{./figures/}}
\usepackage{ams math}
\usepackage{enumerate}
\usepackage{mathbbol}
\usepackage{amsfonts}
\usepackage{natbib}

\usepackage{bm}

\usepackage{color}
\usepackage[dvipsnames]{xcolor}

\usepackage[caption=false]{subfig}
\usepackage{bbm}

\newcommand{\nn}{\nonumber}

\newcommand{\be}{\begin{equation}}
\newcommand{\ee}{\end{equation}}
\newcommand{\bea}{\begin{eqnarray}}
\newcommand{\eea}{\end{eqnarray}}
\newcommand{\beq}{\begin{eqnarray}}
\newcommand{\eeq}{\end{eqnarray}}

\newlength{\bilderlength}

\usepackage{subfig}

\begin{document}

\title{Out of equilibrium dynamics of repulsive ranked diffusions: the expanding crystal}

\author{Ana \surname{Flack}}
\affiliation{LPTMS, CNRS, Univ. Paris-Sud, Universit\'e Paris-Saclay, 91405 Orsay, France}
\author{Pierre Le Doussal}
\affiliation{Laboratoire de Physique de l'Ecole Normale Sup\'erieure, CNRS, ENS and PSL Universit\'e, Sorbonne Universit\'e, Universit\'e Paris Cit\'e,
24 rue Lhomond, 75005 Paris, France}
\author{Satya N. \surname{Majumdar}}
\affiliation{LPTMS, CNRS, Univ. Paris-Sud, Universit\'e Paris-Saclay, 91405 Orsay, France}
\author{Gr\'egory \surname{Schehr}}
\affiliation{Sorbonne Universit\'e, Laboratoire de Physique Th\'eorique et Hautes Energies, CNRS UMR 7589, 4 Place Jussieu, 75252 Paris Cedex 05, France}

\date{\today}

\begin{abstract} 
We study the non-equilibrium Langevin dynamics of $N$ particles in one dimension
with Coulomb repulsive linear %mutual 
interactions. This is a dynamical version of the so-called jellium model (without confinement) also known as ranked diffusion. Using a mapping to the Lieb-Liniger model of quantum bosons, we obtain an exact formula for the joint distribution of the positions of the $N$ particles at time $t$, all starting from the origin. A saddle point analysis shows that the system converges at large time to a linearly expanding crystal. Properly rescaled, this dynamical state resembles 
the equilibrium crystal in a time dependent effective quadratic potential. This analogy allows to study the fluctuations around the perfect crystal,
which, to leading order, are Gaussian. There are however deviations from this Gaussian behavior, which embody long-range correlations of  purely dynamical origin, 
characterized by the higher order cumulants of, e.g., the gaps between the particles, that we calculate exactly. 
We complement these results using a recent approach by one of us in terms of a noisy Burgers equation. 
In the large $N$ limit, the mean density of the gas can be obtained at any time %and for a large class of initial conditions, 
from the solution of a deterministic viscous Burgers equation. This approach provides a quantitative description of the dense regime 
at shorter times. Our predictions are in good agreement with numerical simulations for finite and large $N$. 
%in particular we quantify the deviations at large but finite $N$.
%{\red explain better that equilibrium is amazing}}
\end{abstract}

\maketitle

\tableofcontents

\section{Introduction and the main results}

The Coulomb potential in one dimension is linear in the distance. Particles interacting with this potential have been much studied. 
Many of these studies address the canonical equilibrium at some temperature $T$. In the attractive case it is related to the statistical mechanics of the self-gravitating $1d$ gas \cite{Rybicki,Sire,Kumar2017}. In the repulsive case, and in presence of a background charge or in a finite box, it is called jellium and its fluctuations at equilibrium have been well studied \cite{Lenard,Prager,Baxter,Dean1,Tellez,Lewin}, with a recent renewed interest, in particular in edge fluctuations and large deviations \cite{SatyaJellium1,SatyaJellium2,SatyaJellium3,Chafai_edge,Flack22}.

In this paper, we study the out of equilibrium dynamics of this system and demonstrate that it exhibits rather rich and interesting behaviors, as a function of time. In $1d$, the Coulomb force (either attractive or repulsive) acting on each particle is proportional to its rank, i.e., the number of particles in its front minus the number of particles at the back of it. The Langevin dynamics of this system is called {\it ranked diffusion}. The diffusion of $N$ particles in $1d$ under a drift which depends only on their ranks has been studied in
finance \cite{Banner} and in mathematics \cite{Pitman,OConnell}. 

Recently the non-equilibrium dynamics of this model was studied \cite{PLDRankedDiffusion} 
using a mapping to the Lieb-Liniger model, or $1d$ delta Bose gas. Since the latter is integrable by Bethe ansatz, it allows in principle to obtain formula for non-equilibrium observables in the ranked diffusion model for any $N$.
In practice however this approach is analytically complicated, and not all initial conditions can be easily treated. 
Hence this program has yet to be fully completed. Furthermore the exact solution does not allow to add an external potential, since it breaks integrability. Another more versatile approach was thus also studied in \cite{PLDRankedDiffusion}, which exploits a connection to the noisy Burgers equation. This method is most efficient to study the large $N$ limit, where the effect of the noise term is reduced.

In this paper we focus on the repulsive gas and show that many exact results on its dynamics can be derived %at late times by 
using the aforementioned two complementary approaches. 
We study a gas of $N$ particles on the line, in the absence of external potential, performing thermal diffusion at temperature $T$, 
and mutually interacting via the linear Coulomb potential of strength $c>0$ (see the definition of the model in \eqref{langevin1}). 
In addition to diffusion, each particle thus experiences a drift proportional to its rank, typically of order $O(c\,N)$ for large $N$. 
{There is no additional hard-core interaction and therefore the particles are free to cross each other. 
We focus on the case where all the particles start at $t=0$ from the origin at $x=0$. 
Several realizations of this dynamics are shown in Fig. \ref{Plot_traj} for $N=500$, where one can see that there are several interesting regimes as a function of time. 
Since the gas is expanding from a point source, it is dense at short times and particles experience many mutual crossings (see figure a) in Fig. \ref{Plot_traj}). At large time, the gas is diluted and the particles are far from each other, but, as we will show, they nevertheless form a well ordered expanding crystal due to the long range nature of the interaction (see figure c) in Fig. \ref{Plot_traj}).
In fact we find that when $N$ is large one can distinguish {\it three} different regimes (see figures a),b),c) in Fig. \ref{Plot_traj}). Indeed, there
are two characteristic length scales associated respectively to the diffusion and the drift, namely 
\be \label{length}
\ell_T \sim \sqrt{2 T t} \quad , \quad \ell \sim 2 c N t  \;.
\ee 
The first length $\ell_T$ is the typical thermal diffusion length of independent particles.
Since the rightmost (respectively leftmost) particle experiences a drift $\sim c N$ (respectively $-c N)$
the second length $\ell$ is the total size of the gas at large time. 
Comparing the two length scales we see that there is a characteristic time scale
\be 
t_1^* \sim \frac{1}{N^2} \frac{T}{c^2} 
\ee 
such that for $t < t_1^*$ the diffusion dominates over the drift. In that regime, which we call regime I,
the particles are almost independent and the gas is very dense (see figure a) in Fig. \ref{Plot_traj}).
For time $t > t_1^*$ the drift, i.e., the interaction, dominates over the diffusion. This is regime II,
where the gas evolves from being dense to being dilute. The crossover from regime I to regime II 
in the behavior of the size of the gas (i.e. distance between rightmost and leftmost particle)
is shown in the right upper panel of Fig. \ref{Plot_traj}. 
As we will show, in regime II as time increases, the density converges to a square shape, being
uniform on $[-\ell/2,\ell/2]$, with a boundary layer at the two edges of size $\ell_T \sim \sqrt{2 T t} \ll \ell$.
In this regime II however the particles still experience many crossings, (see panel b in Fig. \ref{Plot_traj})
and the size of the boundary layer is still much larger than the interparticle distance $a= \ell/N = 2 c t$.
As time further increases the density becomes so low that the particles cross each other only rarely.
This happens when $a \sim \ell_T$ which defines the second time scale 
\be 
t_2^* \sim \frac{T}{c^2} 
\ee 
Beyond this time scale, for $t \gg t_2^*$, one has $\ell_T \ll a = 2 c t$, and the particles are well separated and do not cross anymore:
this is regime III (dilute regime). These three regimes, together with the formation of a square density, can be clearly seen in Fig.~\ref{Plot_traj}.
Note that if the initial
condition has instead a finite extension $\ell_0$, there exists another time scale $t^*_0 \sim \ell_0/(c N)$ at which 
most features of the initial density are erased and the plateau forms. Here we mainly focus
on the case $\ell_0=0$ so this time scale is absent. Note that for finite $N$ these time scales are all identical and there
is only a short time regime $c^2 t/T \ll 1$ and a large time regime $c^2 t/T \gg 1$.

\begin{figure}[h!]
\centering
\includegraphics[width=0.85\linewidth]{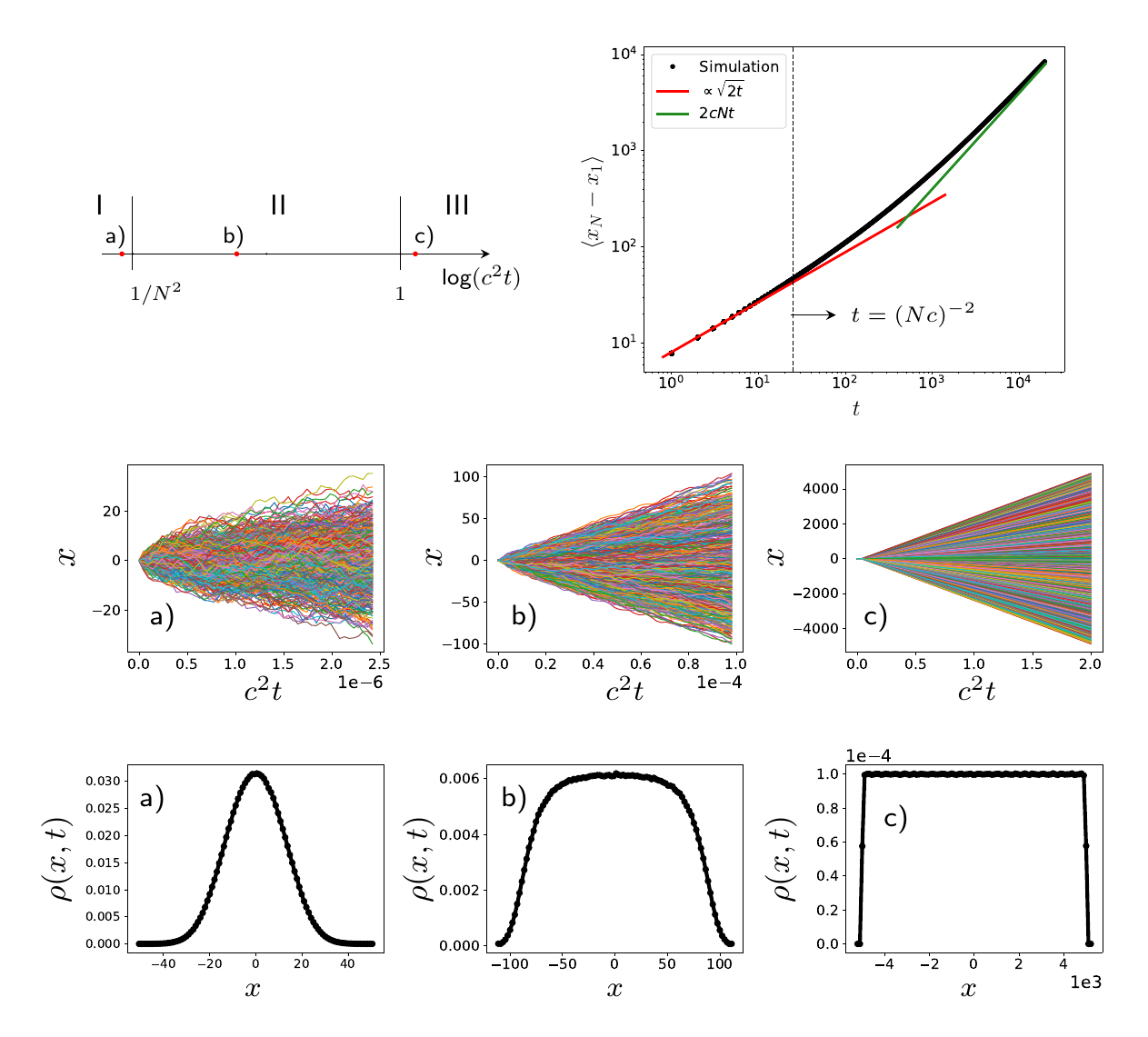}
\caption{ {\it Upper panel left:} summary of the values of the dimensionless final time $c^2t$ used in the figures a), b) and c) below. 
{\it Upper panel right:} size of the gas as a function of time (in a log-log scale) which shows a crossover between the regimes I and II and time $t_1^*=T/(N c)^2$ which is indicated by a vertical dashed line. Symbols are the results of numerical simulations. The red solid line at short time represents the prediction in regime I, $\ell_T = \sqrt{2\,T\, t}$ (here $T=1$), and the green one at larger time corresponds to the prediction 
in regime II, $\ell = 2 N c t$ [see Eq. (\ref{length})]. 
%$2c(N-1)\,t + (2/c)(1-1/N)$ given in Eq. (\ref{span}) with $n=1$. 
{\it Middle panel, figures a-c}: examples of trajectories $x_i(t)$ vs $t$ of $N=500$ particles evolving via the Langevin equation in Eq. (\ref{langevin1}).
One can identify three different regimes determined by the value of the dimensionless time $c^2t$.
{\it Bottom panel, figures a-c}: corresponding densities of particles $\rho(x,t_f)$ at the final time $t=t_f$ for each of the three top figures.
They are obtained by averaging over $10^4$ realizations of the noise, and using $100$ bins to construct the histograms. Here
for convenience we chose $t_f=50$ and varied $c$. These three values fall in each of the three regimes I-III discussed in the text.} 
 \label{Plot_traj}
\end{figure}

%\begin{figure}[t]
%\centering
%\includegraphics[width=0.8\linewidth]{Plots/Unifinshed_plots/Regimes_tr.pdf}
%\caption{The four figures show examples of trajectories in four different regimes %determined by the value $c^2t$. The values of $c^2 t$ are the same as in Fig. %\ref{fig:u3}. The number of particles in simulation is $N=500$.} \label{fig:u4}
%\end{figure}

%\begin{figure}[h]
%\centering
%\includegraphics[width=0.8\linewidth]{Plots/Unifinshed_plots/Regimes_dens.pdf}
%\caption{Density profiles for the four different regimes. The number of particles in %simulation is $N=500$. We averaged over $10^4$ realizations of the noise. The number %of bins is $100$.} \label{fig:u4}
%\end{figure}

These three regimes exhibit quite different density and particle correlation properties. To obtain a quantitative description of the system, we first derive in Section~\ref{sec:model}, using the Bethe ansatz, an integral formula for the joint probability distribution function (PDF) of the positions of the particles, {all starting from the origin}, given in Eq.~\eqref{exactN},
which is exact for any $N$ and $t$. By analyzing this formula via a saddle point method, we obtain in 
Section \ref{sec:three} the asymptotic form of the joint PDF at large time, see Eq. \eqref{rescaled2}. It is a priori
valid for any fixed $N$ and for $c^2 t/T \gg 1$. At large $N$ it thus describes the regime III (the dilute regime). From that formula we
find that in that regime the system is a well ordered expanding crystal, with most probable particle positions $x_j= c (N+1 - 2 j){t}$.
To compute the fluctuations of the particle positions in this crystal, we proceed in two stages. We first approximate the formula \eqref{rescaled2} for the joint PDF
by neglecting the rational prefactor, in which case
it becomes formally identical to the equilibrium distribution given in Eq. \eqref{bolt_ord}
of the $1d$ jellium model [defined in \eqref{E_OCP}]. Although this analogy holds for any $N$, 
it is especially useful for large $N$, where many results are known for the jellium model \cite{SatyaJellium1,SatyaJellium2}. We find that, within that approximation, the regime III corresponds to the jellium model with a dimensionless interaction {strength} $\alpha \gg 1$. 
It correctly predicts the leading order of the fluctuations of the particle positions around their ordered positions, which in this regime
are small and independent Gaussian random variables. The amplitude of these fluctuations are of the order of the diffusion length $\ell_T \sim \sqrt{2 T t}$. 
Next, we treat more accurately the asymptotic joint PDF in Eq. \eqref{rescaled2}, and show that there are non-trivial additional
 position fluctuations, which are of order $O(T/c)$. We characterize them completely by computing
analytically all the joint cumulants of the particle positions, given in Eq. (\ref{cumul1}). These formulae show that there are non-trivial
correlations of purely dynamical origin, which persist for $c^2 t/T \gg 1$ and go beyond the analogy with the equilibrium jellium model. 
%This analysis is valid in the limit where $t \to +\infty$ first, and then $N \gg 1$ in a second stage.
It is yet unclear how to extend these results to the case $c^2 t/T = O(1)$ and in particular to 
the more correlated regime II, i.e., for $ c^2 t/T \ll 1$,
where %the above two types of fluctuations become of the same order. In these regimes 
the particle crossings
cannot be ignored. We expect that some of these effects will be captured by the analogy with the jellium model at finite interaction strength $\alpha$, but that additional dynamical correlations will also exist. Note that we also treat exactly the case $N=2$ which is quite instructive, 
in particular to analyze systematically the role of the initial condition (which we were not able to do for general $N$). 

Next, in Section \ref{sec:burgers}, we recall the hydrodynamic approach of \cite{PLDRankedDiffusion}
using the Burgers equation, 
which gives a prediction for the time dependent density at large $N$ for arbitrary time. 
%It thus allows to handle the opposite limit $N \to +\infty$ first, and then $t \gg 1$, and thus describes
It thus allows to derive analytical formula for the time evolution of the average particle density within the crossover between the regimes I and II, when
the gas is still sufficiently dense.
%(see left panel in Fig. \ref{Plot_traj}), 
For that study it is convenient to scale $c=\gamma/N$ with $\gamma=O(1)$
in which case the time scale $t_1=O(1)$. Finally, we perform
numerical simulations and compare the results with the predictions of both methods (the Bethe ansatz and the hydrodynamic approach). 
We confirm the Gaussian character of the fluctuations in regime III and we
test the accuracy of the predictions for the density using the deterministic Burgers equation.

Finally, in Appendix A we study in detail the case $N=2$. In Appendix B we study the corrections to the cumulants of the particle
positions, which are exponentially small at large time. In Appendix C we derive the form of the boundary layer for the Burgers equation.

% \begin{figure}[h]
% \includegraphics[width = 0.8 \linewidth]{Plots/Trajectories_B.pdf}
% \caption{The trajectories $x_i(t)$ vs $t$ of $N=50$ particles evolving via the Langevin equation in Eq. (\ref{langevin1}) 
% with $T=1$ and all starting from the origin at $t=0$ for two different values of $c$. Note that the dimensionless time parameter is $c^2 t$.
% {\red P: should we plot it instead versus $c^2 t$} }\label{Plot_traj}
% \end{figure}

\section{Model and main formula} \label{sec:model}

In this paper we consider $N$ particles on the real line at positions $x_i(t)$, $i=1,\dots,N$, evolving according to the Langevin equation
\bea \label{langevin1}
\frac{dx_i}{dt} = - \partial_{x_i} W(\vec x) + \sqrt{2 T} \xi_i(t)  = - c \sum_{j=1}^N {\rm sgn}(x_j-x_i) + \sqrt{2 T} \xi_i(t) \;,
\eea 
where $\xi_i(t)$ are $N$ unit independent white noises with zero mean $\langle \zeta_i(t)\rangle = 0$ and delta-correlator $\langle \zeta_i(t) \zeta_j(t')\rangle = \delta_{i,j}\delta(t-t')$. Here $T$ is the temperature, and by convention ${\rm sgn}(0)=0$. 
The particles interact via the linear pairwise potential energy $W(\vec x) = - c \sum_{i<j} |x_i-x_j|$
where we denote ${\vec x}= \{ x_i(t) \}_{i=1,\dots,N}$.
The particles may cross (and they will) and if we denote $x_{(i)}(t)$ the ordered sequence
of their positions at time $t$ in increasing order, then the ordered particle $x_{(i)}$ feels a drift 
\bea \label{drift}
\delta_i = - c \sum_{j=1}^N {\rm sgn}(x_{(j)}-x_{(i)}) = - c\,(N+1- 2 i) \;,
\eea 
which depends on the label/rank $i$ of the particle: this is just proportional to the number
of particles in front minus the number of particles at the back of the $i$-th particle. 
For $c>0$ the interaction is thus repulsive, the case considered here.

Next one introduces the probability density function (PDF), $P(\vec x,t)$, of a given configuration
${\vec x}$ of the particles. It satisfies the 
Fokker-Planck (FP) equation 
\be \label{FP0} 
\partial_t P = - {\cal H}_{\rm FP} P =  \sum_i [ T \partial_{x_i}^2 
- c \partial_{x_i}   \sum_j {\rm sgn}(x_i-x_j)  ] P \;.
\ee 
For $c<0$ this equation formally admits a zero current stationary solution 
\bea
P_0(\vec x) = \frac{1}{Z_N} \Psi_0(\vec x)^2  \quad , \quad \Psi_0(\vec x) : = e^{-\frac{1}{2 T} W(\vec x)} =
e^{\frac{c}{4 T} \sum_{i,j=1}^N |x_i-x_j| }  \;,
\eea
where $Z_N$ is a normalization constant. For $c>0$ this solution is however not normalizable, and is not the stationary state. Indeed, in the absence of external potential the 
gas expands linearly with time \cite{PLDRankedDiffusion}, an expansion that we will study here in more details.

It is useful to note at this stage that the two parameters of the model, $T$ and $c$, can
be absorbed in a change of units. More precisely $T/c$ is a length scale and $T/c^2$ is a time scale. 
In terms of these scales one can always write
\be \label{Ptilde}
P(\vec{x},t) = {\left(\frac{c}{T}\right)^N} \tilde P\left(\frac{c \, \vec{x}}{T} , \frac{c^2 t}{T} \right) \;,
\ee 
where $\tilde P(\vec x,t)$ is the PDF for the model with $c=T=1$. We have seen in the introduction
that at large $N$ there are several distinct time scales. These can be explored conveniently by
scaling $c$ in various ways with $N$. {Hence we will not fix the parameter $c$. However for the calculations in the remainder of this section, 
as well as in Section \ref{sec:three}, we will set $T=1$. Since $c$ and $T$ can be absorbed in the
units, there is no intrinsic dimensionless parameter in the model, besides $N$ and some parameter characterizing the
initial condition, e.g such as $c {\ell}_0/T$, if ${\ell}_0$ is the initial interparticle distance (below we 
focus on $\ell_0=0$). Hence all the regimes can be obtained by looking at the
particular scale of interest.}

Let us consider now the delta initial condition where all particles are at the same
position $\vec x(0)=\vec 0$ in space at time $t=0$
\be \label{delta} 
P(\vec x,t=0) = \prod_i \delta(x_i) \;.
\ee 
Hence $P(\vec x, t)$ is the Green's function of the Fokker-Planck operator ${\cal H}_{\rm FP}$, i.e., 
$P(\vec x, t) = G_{\rm FP}(\vec x, \vec 0,t)$ where 
$G_{\rm FP}(\vec x, \vec y,t) = \langle \vec x |e^{- t {\cal H}_{\rm FP}}|\vec y\rangle$. It is easy to check that 
\bea  \label{rel_Green}
G_{\rm FP}(\vec x, \vec y,t) =  \frac{\Psi_0(\vec x)}{\Psi_0(\vec y)} G_{s}(\vec x, \vec y,t) e^{E_0\,t} \quad {\rm with} \quad E_0 = -\frac{c^2}{12}(N^3-N) \;,
\eea 
where $G_{s}(\vec x, \vec y,t)$ is the Green's function of the Schr\"odinger Hamiltonian ${\cal H}_s$,
i.e., the solution for $t>0$ of $\partial_t G_s = - {\cal H}_{s} G$ with initial condition
$G_{s}(\vec x, \vec y,t=0)= \prod_i \delta(x_i)$. Here ${\cal H}_s$ is the 
Lieb-Liniger Hamiltonian \cite{LL}
\be \label{LLmodel} 
 {\cal H}_s = 
- \sum_i \partial_{x_i}^2  + 2 c \sum_{1 \leq i < j \leq N} \delta(x_i-x_j) 
\ee 
which describes quantum particles with delta repulsive interactions. Note that
the initial condition \eqref{delta} is symmetric in the exchange of particles.
In this symmetric sector the model \eqref{LLmodel} is also called the delta Bose gas, which is integrable by the Bethe ansatz
\cite{LL,GaudinBook}. The quantity $E_0$ is the ground state energy of the model. 
The relation \eqref{rel_Green} can be checked by applying ${\cal H}_{\rm FP}$ on each side.

% Nevertheless we can still perform the change of function (from now on we set $T=1$).
% \be  \label{change} 
% P(\vec x,t) = \Psi_0(\vec x) \Psi(\vec x, t) e^{E_0 t} \quad , \quad E_0 = - \frac{c^2}{12} (N^3-N)
% \ee 
% and $\Psi(\vec x,t)$ then satisfies 
% \be \label{LLmodel} 
% \partial_t \Psi(\vec x,t) = - {\cal H}_s \Psi(\vec x,t) \quad , \quad {\cal H}_s = 
% - \sum_i \partial_{x_i}^2  + 2 c \sum_{1 \leq i < j \leq N} \delta(x_i-x_j) 
% \ee 
% where ${\cal H}_s$ is the Lieb-Liniger Hamiltonian \cite{LL}
% describing quantum particles with delta repulsive interactions. 
% \\
% We now use a known result for the Lieb-Liniger model. Let us consider the delta initial condition where all particles are at the same
% position in space at time $t=0$
% \be \label{delta} 
% P(\vec x,t=0) = \prod_i \delta(x_i) 
% \ee 
% It implies the same initial condition for $\Psi(\vec x,t=0)= \prod_i \delta(x_i)$. This initial condition
% is symmetric in the $\{x_i\}$, hence its evolution involves only symmetric (i.e. bosonic) eigenstates of \eqref{LLmodel}.
% Restricted to this symmetric sector, \eqref{LLmodel} is called the delta Bose gas, which is integrable by the Bethe ansatz
% \cite{LL,GaudinBook}. 

For the delta initial condition the Schrodinger problem can be solved and 
there exists a multiple integral formula for $G_s(\vec x,\vec 0,t)$ valid for all times.
From Proposition 6.2.3 (Eq. (6.6)) in \cite{BorodinCorwinMacDo} (for earlier works see 
\cite{TWBoseGas,GaudinBook}), and using \eqref{rel_Green} 
we obtain for $c>0$, for any $N$ and $t>0$ and for the sector $x_1 \leq x_2 \leq \dots \leq x_N$
% \be \label{exactN} 
% P(\vec x,t) = e^{\frac{c}{4} \sum_{i,j=1}^N |x_i-x_j| }  e^{E_0 t} 
% \int_{i \mathbb{R}} \frac{dz_1}{2 i \pi} \dots \int_{i \mathbb{R}}  
% \frac{dz_N}{2 i \pi} \prod_{1 \leq a < b \leq N} \frac{z_a-z_b}{z_a - z_b + c} \quad e^{t \sum_{j=1}^N z_j^2 + \sum_{j=1}^N  x_j z_j} 
% \ee 
\be \label{exactN} 
{P(\vec x,t) = G_{\rm FP}(\vec x, \vec y = \vec 0,t) =} e^{\frac{c}{4} \sum_{i,j=1}^N |x_i-x_j| }  e^{E_0 t} 
\int_{\mathbb{R}} \frac{dk_1}{2 \pi} \dots \int_{\mathbb{R}}  
\frac{dk_N}{2 \pi} \prod_{1 \leq a < b \leq N} \frac{i k_a- i k_b}{i k_a - i k_b + c} \quad e^{- t \sum_{j=1}^N k_j^2 + i \sum_{j=1}^N  x_j k_j} \;,
\ee 
where the variables $k_i$ are integrated over the real axis and $E_0$ is given in Eq. (\ref{rel_Green}). Being a fully symmetric function of its arguments, $P(\vec x , t)$ is obtained
in the other sectors by symmetry. Note that, by construction, $P(\vec x,t)$ in Eq. (\ref{exactN}) is normalized to unity on $\mathbb{R}^N$ for 
%i.e. $\int dx_1 \cdots dx_N P(\vec x,t) = 1$ for 
all time $t \geq 0$, although this property is not so obvious to check from (\ref{exactN}). This explicit expression in Eq. (\ref{exactN}) is our main formula, which we analyze in the following sections.

\section{Time evolution of $P(\vec{x},t)$} \label{sec:three}

In this section, we analyze the time-evolution of $P(\vec{x},t)$. In the first subsection \ref{subsec:N2}, we verify the validity of the formula (\ref{exactN}) for $N=2$ by finding directly the exact solution of the Fokker-Planck equation (\ref{FP0}). {The case $N=2$ being already very instructive, we study in
details its large time asymptotics.} In the next subsection \ref{subsec:saddle}, we perform a saddle-point analysis of the formula in (\ref{exactN}) for large $t$, for any fixed $N$. {In subsection \ref{subsec:equilibrium}, we make an analogy between this expanding Coulomb gas at large $t$ and the static properties of a one-dimensional one-component plasma in a harmonic potential. Finally, in subsection \ref{subsec:cumulants} we obtain the higher cumulants of the position fluctuations from a more precise analysis of the large time limit.}

\subsection{Two particles $N=2$} 
\label{subsec:N2} 

Let us start with two particles, i.e., $N=2$. Introducing the center of mass coordinate
$x(t)=\frac{1}{2} (x_1(t) + x_2(t))$ and the relative coordinate $y(t)=x_2(t)-x_1(t)$,
the Fokker Planck equation \eqref{FP0} 
is easily solved directly by a Laplace transform. 
{Denoting $P(y,t)$ the PDF of $y(t)$ (with a slight abuse of notation) we find} (see details in Appendix \ref{app:twoparticles}) 
\be  \label{Laplace2} 
\tilde P(y,s) = \int_0^{+\infty} dt e^{-s t} P(y,t) = \frac{e^{\frac{c}{2} |y|} }{2(c + \sqrt{c^2 + 2 s})} e^{- \frac{1}{2} \sqrt{c^2+ 2 s} |y|} \;,
\ee 
which upon Laplace inversion gives
\be \label{result2} 
P(y,t)= \frac{e^{-\frac{(|y|-2 c t)^2}{8 t}}}{2 \sqrt{2 \pi }
   \sqrt{t}}-\frac{1}{4} c e^{c |y|}
   \text{erfc}\left(\frac{2 c t+|y|}{2 \sqrt{2}
   \sqrt{t}}\right) \;,
\ee 
{which is normalized to unity, i.e., $\int_{-\infty}^{+\infty} dy P(y,t)=1$ and satisfies the initial condition $P(y,0) = \delta(y)$}. Here ${\rm erfc(z)} = 2/\sqrt{\pi}\int_z^{\infty}e^{-u^2}\,du$. We now want to check that the general formula in Eq. (\ref{exactN}) for the joint distribution of $N$ particles also leads to the result in Eq. (\ref{result2}) for $N=2$. Indeed, in Appendix \ref{app:twoparticles}, we show this explicitly. The time evolution of $P(y,t)$ from \eqref{result2} is plotted in Fig. \ref{N2}.

\begin{figure}[t]
    \centering
    \includegraphics[width=0.8\textwidth]{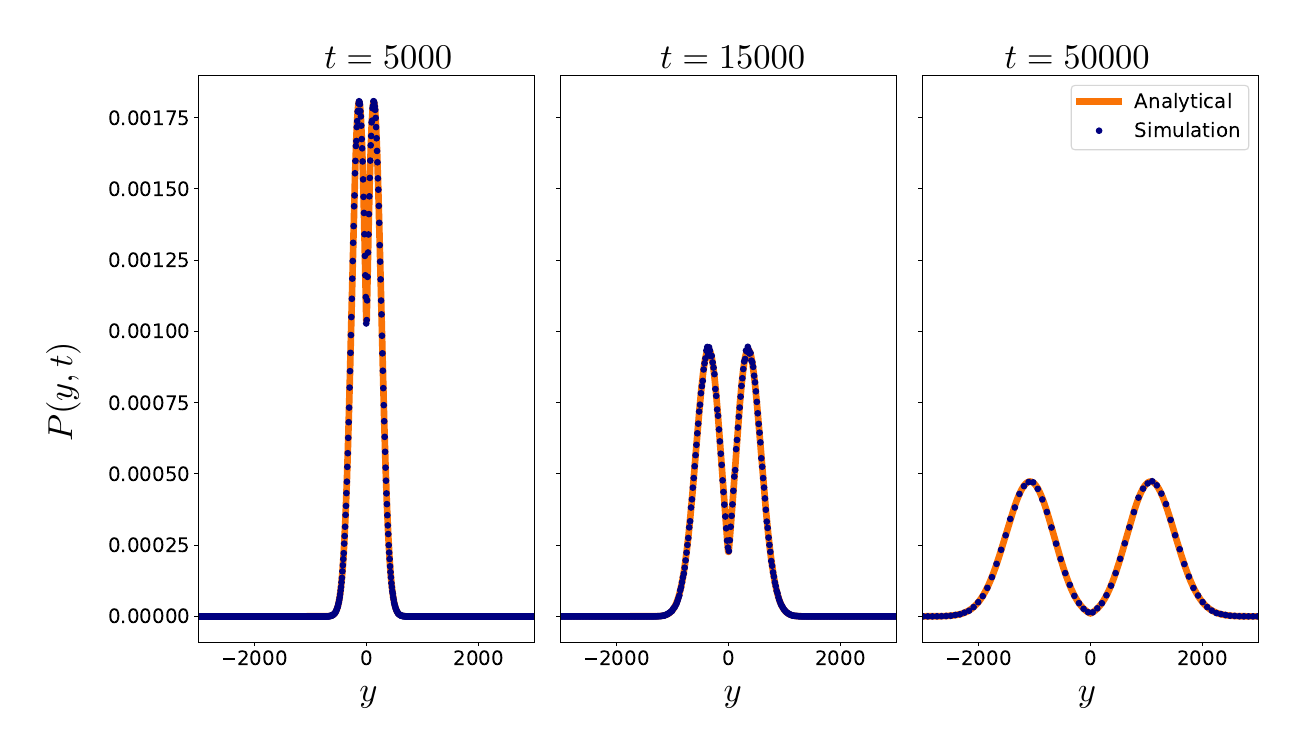}
    \caption{The probability distribution $P(y,t)$ of the relative coordinate $y(t)=x_2(t)-x_1(t)$ for two particles for different times ($t=5000$, $t=15000$, $t=50000$). The blue dots are obtained by the simulation with $c=0.01$, $T=1$ and averaging over $10^6$ realizations. The solid orange lines represent the analytical expression for the distribution derived in Eq. \eqref{result2}. One can see that the distribution becomes bimodal at large time.}\label{N2}
\end{figure}

At large time it becomes a bimodal distribution centered around $y \simeq \pm 2 c t$. For fixed $y = O(1)$ one finds, as $t \to \infty$
% {\red P: I have rechecked all normalizations and limits. There was a subtelty but it is resolved, see the 
% commented text in the tex file}
% {\red P: there is something funny with \eqref{yfixed}. On one hand I have checked numerically that \eqref{result2} 
% is normalized to unity for $y \in \mathbb{R}$ (both sides), and \eqref{yfixed} is the asymptotics that mathematica gives me.
% But then when I compute the normalization of \eqref{yfixed} I find that it goes to 2 at large $t$. On the
% other hand I find that the normalization $\int_\mathbb{R} dy = t \int_\mathbb{R} dz$  
% of \eqref{N_2_z}  is $1$ as it should. And the two formula match, as I have added in blue. So there is a mystery.}
% Mystery solved, this formula is valid only for $y=O(1)$ which is not where the normalization lies. Computing
%the normalization this way is thus incorrect, it is just a coincidence that it gives $2$ (this factor
%can in fact be understood from the matching
\be \label{yfixed} 
P(y,t) \simeq \frac{(c |y|+2) e^{-\frac{(|y|-2 c t)^2}{8 t}}}{4 \sqrt{2 \pi } c^2 t^{3/2}} { \simeq
\frac{(c |y|+2) e^{\frac{|y|}{2}} }{4 \sqrt{2 \pi } c^2 t^{3/2}} e^{- \frac{c^2 t}{2}} }
\;.
\ee
where in the last equation we used that $|y|=O(1)$.
On the other hand, if one scales $y = z t$ with fixed $z=O(1)$ one finds
\be 
P(y,t) \simeq \frac{ |z| e^{-\frac{1}{8} t (|z|-2 c)^2}}{2 \sqrt{2 \pi t } \label{N_2_z}
    (|z| + 2 c ) } \;.
\ee 
On these scales it thus converges, as $t \to \infty$, to a pair of delta-functions {at $z=\pm 2$, each of weight $1/2$.
For $t$ large but finite one can check that the total probability weight in the asymptotic form \eqref{N_2_z} is slightly
less than unity, but converges to unity as $t \to +\infty$. 
%{\red P: it converges as a power law in $1/t$, while the exact cumulants seem to converge exponentially, which is strange}.
Note that although the exponential factor is the same in \eqref{yfixed} and in \eqref{N_2_z},
the prefactors in each formula are distinct: only the large $|y|=O(1)$ limit of \eqref{yfixed} 
matches the small $|z|=O(1)$ limit of~\eqref{N_2_z}.} 

One can also calculate the cumulants of the random variable $|y|- 2 c t$. While more detailed expressions are
given in Appendix \ref{sec:app_cumul}, here we simply indicate their leading behaviors at large time. We obtain
\bea \label{meanN2} 
&& \langle |y| \rangle - 2 c t  \simeq 1/c + O(t^{-3/2} e^{-c^2 t/2}) \;, \\
&& \langle (|y|-2 c t)^2 \rangle_c  \simeq 4 t - 3/c^2 + O(t^{-1/2} e^{- c^2 t/2})\;, \label{var_N2}\\
&& \langle (|y|-2 c t)^3 \rangle_c  \simeq 14/c^3 + O(t^{1/2} e^{-c^2 t/2})\;, \\
&& \langle (|y|-2 c t)^4 \rangle_c  \simeq -90/c^4 + O(t^{3/2} e^{-c^2 t/2}) \;.  \label{cum4N2} 
\eea 
where $\langle \dots \rangle_c$, with a subscript $c$, denotes cumulants (not to be confused with the interaction parameter $c$).
Interestingly, while the variance of $|y|$  grows linearly with $t$ for large $t$, the higher cumulants converge to a constant.  
The leading fluctuations are thus Gaussian and diffusive $O(\sqrt{t})$, but there are some additional $O(1)$ non Gaussian fluctuations, as encoded in the higher order cumulants. {As we will see below these features will extend to any $N$, hence $N=2$ is a useful testing ground for general $N$.
One can check (e.g. numerically) that the leading orders (i.e., up to $O(1)$) of the cumulants \eqref{meanN2}-\eqref{cum4N2} 
are reproduced if one uses the asymptotic form \eqref{N_2_z}, which thus captures the $O(1)$ non Gaussian fluctuations at large time.
In addition, in Section \ref{subsec:cumulants} we obtain an {\it exact formula} for the $O(1)$ leading orders of all the cumulants at large time, i.e,
for any $N$ (hence including $N=2$), based on a saddle point method. To obtain them we show that for $c^2 t \gg 1$ one can restrict to
an ordered sector $x_1 < \dots < x_N$ and neglect the events when particles
cross. These events are only responsible for the exponential corrections to the cumulants (e.g. $e^{-c^2 t/2}$ for
$N=2$ in \eqref{meanN2}-\eqref{cum4N2}). Finally, we have checked the predictions of \eqref{meanN2}-\eqref{cum4N2} by a numerical solution of the
Langevin equation, the results are presented in Fig. \ref{fig:u2}.
}

\begin{figure}[h!]
\centering
\includegraphics[width=\linewidth]{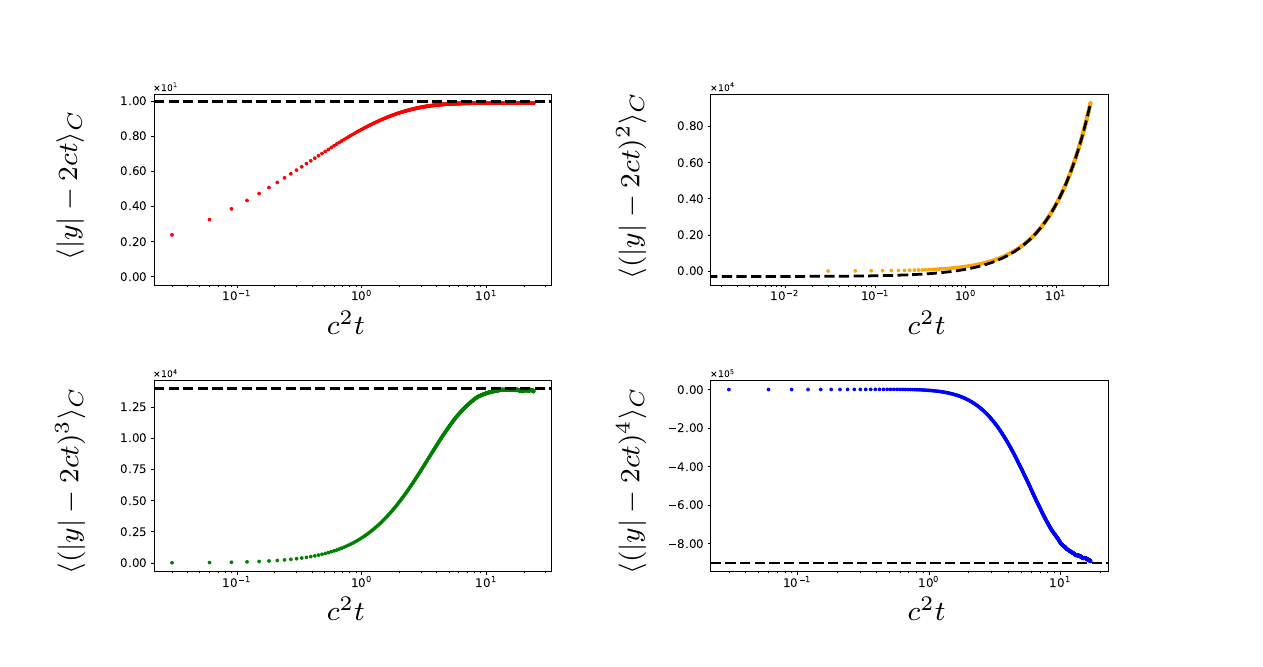}
\caption{Plot of first four cumulants $\langle (|y| -2 c t)^k \rangle_c$, with $k=1,2,3,4$ as a function of dimensionless time $c^2 t$ (in logarithmic scale) from 
the numerical solution of the Langevin equation, as 
compared to the analytical prediction at large time from Eqs. \eqref{meanN2}-\eqref{cum4N2} (black dashed line). 
We chose $c=0.1$ and averaged over $2\times 10^8$ realizations of the noise.
} \label{fig:u2}
\end{figure}

\subsection{Saddle-point analysis at late times $t$ for fixed $N$} 
\label{subsec:saddle} 

Let us now consider the general case of $N$ particles. We start from the general formula \eqref{exactN} for the joint PDF restricted to the sector
$x_1 \leq x_2 \leq \dots \leq x_N$. Let us define the rescaled variables $x_j = z_j t$. In terms of these variables the 
joint PDF $P(\vec x,t)$ reads, using $E_0$ from Eq. (\ref{rel_Green}) and with $z_1 \leq z_2 \leq \dots \leq z_N$,
% \be \label{exactN} 
% P(\vec x,t) = e^{\frac{c}{4} \sum_{i,j=1}^N |x_i-x_j| }  e^{E_0 t} 
% \int_{\mathbb{R}} \frac{dk_1}{2 \pi} \dots \int_{\mathbb{R}}  
% \frac{dk_N}{2 \pi} \prod_{1 \leq a < b \leq N} \frac{i k_a- i k_b}{i k_a - i k_b + c} \quad e^{- t \sum_{j=1}^N k_j^2 + i \sum_{j=1}^N  x_j k_j} 
% \ee 
\bea  \label{rescaled0} 
P(\vec x,t) = \int_{\mathbb{R}} \frac{dk_1}{2 \pi} \dots \int_{\mathbb{R}}  
\frac{dk_N}{2 \pi}  
e^{t\,[\frac{c}{4} \sum_{i,j=1}^N |z_i-z_j| - \sum_{j=1}^N k_j^2 + i \sum_{j=1}^N  z_j k_j  - \frac{c^2}{12} N (N^2-1)  ]} 
\prod_{1 \leq a < b \leq N} \frac{i k_a- i k_b}{i k_a - i k_b + c} \;.
\eea 
Let us consider now the regime of large time, $t \to +\infty$ with fixed $N$, and with the $z_j = O(1)$, i.e., $x_j = O(t)$. For large $t$, the expression multiplying $t$ in the exponent of the integrand in Eq. (\ref{rescaled0}) gets minimized at the saddle point with the values
\be 
k_j^* = \frac{i z_j}{2} \;,
\ee 
which are on the imaginary $k$-axis. Note that the original integrals in (\ref{rescaled0}) are on the real $k$-axis. Hence we need to deform the contour in the complex $k$-plane so that it passes through the saddle point and picks up the leading contribution for large $t$. {This can be done without crossing the poles in the prefactor in (\ref{rescaled0}), by deforming the contours for each $k_j$ successively maintaining the condition ${\rm Im} \, k_N > \cdots > {\rm Im} \, k_1$. This gives}  
\bea  \label{rescaled} 
P(\vec x,t) \simeq \frac{1}{(4 \pi t)^{N/2}} e^{\frac{t c}{4} \sum_{i,j=1}^N |z_i-z_j| - t \sum_j \frac{z_j^2}{4} }  
e^{- \frac{c^2}{12} N (N^2-1)  t}  
\prod_{1 \leq a < b \leq N} \frac{z_b-z_a}{z_b - z_a + 2 c}  \;.
\eea 
Note that since $z_b>z_a$ for $b>a$ there are no poles in the double product. For $N=2$ one can 
check that one recovers the expression in \eqref{N_2_z}, which is a bimodal distribution at large time.
{As was discussed there, it is valid for $c^2 t \gg 1$, and so is \eqref{rescaled} for any finite $N$.}

One can rewrite this formula to make more explicit the most probable position of each particle. 
Using the equality, for $z_1 \leq z_2 \dots \leq z_N$ 
\be \label{id1}
\frac{c}{4} \sum_{i,j=1}^N |z_i-z_j| = - \frac{c}{2} \sum_{j=1}^N (N+1-2 j) z_j
\ee 
and completing the square, one finds 
\bea  \label{rescaled2} 
P(\vec x,t) \simeq \frac{1}{(4 \pi t)^{N/2}} e^{- \frac{t}{4} \sum_{j=1}^N (z_j - c (2 j - N - 1))^2  }
\prod_{1 \leq a < b \leq N} \frac{z_b-z_a}{z_b - z_a + 2 c} 
\quad , \quad \text{for} \, \, z_1 \leq z_2 \dots \leq z_N  \;.
\eea 
The most probable values for the rescaled positions at large time are thus 
\be \label{eq_xj}
\frac{x_j}{t}= z_j = c (2 j - N - 1) \;.
\ee 
% Can one check that this is normalized to unity ? In the variables $z_j$ this becomes for any $N$
% a Dirac comb with $z_j$ given by minimization of 
% \be 
% z_j = {\rm argmin} [ - c \sum_{i,j=1}^N |z_i-z_j| +  \sum_{j=1}^N z_j^2 ]  = c (2 j - N - 1) 
% \ee 
These positions form a perfect crystal with uniform spacing $2 c$ which extends from $z_1 = - c (N-1)$ to $z_N = c (N-1)$. 
One can check that for $t \to +\infty$ the normalization of the formula \eqref{rescaled2} inside the ordered sector
is $1/N!$, as expected. Indeed in that limit (i) the fluctuations of the $z_j$'s around the most probable values are vanishing as $1/\sqrt{t}$
(ii) in the prefactor one can simply replace the $z_a$ by their most probable values, 
$z_a=c (2 j - N - 1)$ and one finds that the double product over $a$ and $b$ in (\ref{rescaled2}) simply equals $1/N!$. 
\\

We will now, and in the following subsections, ask about the deviations around the perfect crystal.
Let us denote them as 
\bea \label{def_deltax}
\delta x_j = x_j - c \,t (2 j - N - 1) %{\quad}, \quad {\rm and} \quad \delta z_j=\delta x_j/t \;.
\eea
The quadratic form in the exponential in \eqref{rescaled2} can be rewritten simply as $- \sum_j \frac{\delta x_j^2}{4 t}$. 
This would suggest that the fluctuations of are the $\delta x_j$'s are independent and Gaussian for each particle
with a width given by the diffusion length, $\ell_T = \sqrt{2 t}$, independently of $N$. This is
{\it not the case} however for the two following reasons: 
\begin{itemize}
\item[(i)] there is an ordering condition between the particles,  

\item[(ii)] there is the double product prefactor in \eqref{rescaled2}. 
\end{itemize}

Nevertheless, it is true that if one scales $\delta x_j = \sqrt{2 t} \, \delta \tilde x_j $ with $\delta \tilde x_j = O(1)$
the joint PDF of the $\delta \tilde x_j$'s converges, as $c^2 t \to +\infty$, to a product of independent standard Gaussian variables.
Indeed, with that scaling, the crossing events have an exponentially small probability of order $O(e^{- c^2 t/2})$
(as estimated by displacing two neighbors by $\delta x_j= c t$, $\delta x_{j+1}=- c t$). Hence the neglect of (i)
is justified with that scaling. In addition neglecting the fluctuations $\delta \tilde x_j/\sqrt{t}$
of the variables $z_j=x_j/t$ in the prefactor in \eqref{rescaled2} is also legitimate with that scaling.

Returning to the {\it unscaled} displacements, $\delta x_j$, we have already seen for $N=2$ that their cumulants have 
non trivial additional $O(1)$ contributions, plus exponential corrections of $O(e^{- c^2 t/2})$, see Eqs. \eqref{meanN2}-\eqref{cum4N2}. 
Thus there are interesting deviations due to (i) and (ii) to the independent Gaussian picture. 
We will discuss them in the next two subsections, first neglecting (ii), which
leads to an analogy with an equilibrium problem, and second performing a more accurate analysis of (ii). 

The above considerations are exact at large time $c^2 t \gg 1$ for any $N$. For large $N$ this corresponds to
the regime III as defined in the Introduction.

% \subsection{large $N$ any time {\red P: in construction}}

% we want to scale all terms in the same way $x_j \sim \ell$, $k_j \sim b$ one must have 
% \be 
% c N^2 \ell \sim t N b^2 \sim b \ell N \sim \sum_{a<b} \log( \frac{i k_a-i k_b}{i k_a - i k_b + c} ) 
% \ee 
% This already implies
% \be 
% c=b/N \quad , \quad t \sim \ell/b 
% \ee 
% The last term is more problematic. Assume $c \ll b$. Then 
% \be 
% \sum_{a<b} \log( \frac{i k_a-i k_b}{i k_a - i k_b + c} ) 
% \simeq \sum_{a<b}  \frac{- c}{i k_a - i k_b} 
% \ee 
% Let us assume that this is of order $c N^2/b$, which from the above is $N$. Hence we have
% \be 
% c=b/N \quad , \quad t \sim \ell/b \sim 1/b^2  \quad , \quad \ell \sim 1/b
% \ee 
% If we choose $c=1$ this gives
% \be 
% c=1 \quad , \quad b=N  \quad , \quad t \sim 1/N^2  \quad , \quad \ell \sim 1/N
% \ee 

% The saddle point equations are {\red recheck} 
% \be 
% \frac{c}{2} \sum_\ell {\rm sgn}(x_j-x_\ell)  + i k_j = 0 
% \quad, \quad -2 t k_j + i x_j - c \sum_{a > j} \frac{1}{(i k_j - i k_a)^2} + c  
% \sum_{a < j} \frac{1}{(i k_j - i k_a)^2} = 0 
% \ee 

\subsection{{Analogy with the equilibrium one-dimensional one-component plasma (jellium)}}
\label{subsec:equilibrium} 

In this subsection, we want to make a {comparison} %connection 
between the time-dependent problem of ranked diffusion, 
characterized by $P(\vec{x},t)$ in Eq. (\ref{rescaled2}) and the equilibrium problem of the jellium model in one-dimension (variantly
called the one-dimensional one-component plasma). The jellium model in one-dimension consists of $N$ particles confined in a harmonic
potential and repelling each other via a pairwise Coulomb interaction (which is linear in $1d$). The energy function can be written as \cite{SatyaJellium2,SatyaJellium3}
\be  \label{E_OCP}
E[\vec{y}] = \frac{N^2}{2} \sum_i y_i^2 - \alpha N \sum_{i \neq j} |y_i-y_j| \;,
\ee
where $y_i$'s are assumed to be of order $O(1)$. The first term describes the potential energy, while the second term describes the interaction
energy. {Here $\alpha$ is the strength of the interaction, and is a dimensionless parameter.} The system is supposed to be at equilibrium at temperature $T_{\rm eq}$ and the stationary probability distribution of the positions of the particles is given by the Gibbs-Boltzmann form
\bea \label{boltz}
P_J[\vec{y}] = \frac{1}{Z_N} e^{-E[\vec{y}]/(k_B T_{\rm eq})} \;,
\eea
where $k_B$ is the Boltzmann constant and $Z_N$ is the normalizing partition function. In Eq. (\ref{boltz}), the subscript '$J$' refers to the jellium model. We henceforth set $k_B T_{\rm eq} = 1$ for convenience. It turns out to be convenient to re-write the energy in Eq. (\ref{E_OCP}) in terms of the ordered coordinates $y_1< y_2<\cdots< y_N$. In terms of these ordered coordinates, using the identity in Eq. (\ref{id1}), we get
\bea \label{ord_E_OCP}
E[\vec{y}] = \frac{N^2}{2} \sum_{i=1}^N \left[ y_i - \frac{2\alpha}{N}(2i-N-1) \right]^2 - C_N(\alpha)  \;,
\eea  
where the constant $C_N(\alpha)$ is given by
\bea \label{CN}
C_N(\alpha) =  2 \alpha^2 \sum_{i=1}^N (2i-N-1)^2 =  \frac{2\alpha^2}{3}N^3- \frac{2}{3}\,\alpha^2\,N \;. 
\eea
This implies that, in the ordered sector, the probability distribution of the $y_i$'s can be written as 
\bea \label{bolt_ord}
P_J[\vec{y}] = \frac{1}{\tilde Z_N} e^{-\frac{N^2}{2}\sum_{i=1}^N \left[ y_i - \frac{2\alpha}{N}(2i-N-1)\right]^2} \quad, \quad {\rm for} \quad y_1<y_2<\cdots< y_N\;,
\eea
where $\tilde Z_N$ is a normalization constant. The distribution has a maximum when $y_i$'s occupy the equidistant 
crystal positions, i.e.,
\bea \label{y_crystal}
y_i^* = \frac{2\alpha}{N}(2i-N-1) \;.
\eea
The separation between successive particles is thus $4 \alpha/N$. Defining the equilibrium density (normalised to unity) as 
%{\red P: maybe call it $\rho_{eq}$ or $\rho_J$ to distinguish it from the density in the Burgers section}
\bea \label{def_rho}
\rho_J(y)  = \frac{1}{N} \sum_{i=1}^N \langle \delta(y-y_i) \rangle \;,
\eea
where $\langle \cdots \rangle$ denotes an average over the equilibrium measure in Eq. (\ref{bolt_ord}). Using Eq. (\ref{y_crystal}), we see that in the $1d$-jellium model at equilibrium, the density in the large $N$ limit converges to a flat distribution supported over $[-2\alpha, +2\alpha]$, i.e.,
\bea \label{rho_lim}
\rho_J(y) \approx \frac{1}{4 \alpha} \mathbb{I}_{[-2\alpha, +2 \alpha]}(y) \;,
\eea
where the indicator function $\mathbb{I}_{[-2\alpha, +2 \alpha]}(y)$ is $1$ for $y \in [-2\alpha, +2 \alpha]$ and is $0$ outside. From Eq. (\ref{bolt_ord}), one can also infer the statistics of the positions of the particles in the gas in the two opposite limits: (i) noninteracting limit $\alpha \to 0$ and (ii) the strongly interacting limit $\alpha \to \infty$. In case (i), the particles are essentially independent, each having Gaussian fluctuations around the origin, with width $1/N$. In case (ii) the particles are localized at the crystal positions in Eq. (\ref{y_crystal}), namely $y_i^* = \frac{2\alpha}{N}(2i-N-1)$ and around each position, the fluctuations are again Gaussian {and independent} with width $1/N$. The crossover between the two cases occurs for $\alpha=O(1)$, {where
the correlations are non trivial.}

In order to compare this equilibrium problem with the dynamics of ranked diffusion discussed earlier, 
{we consider the asymptotic form of the probability distribution $P(\vec x,t)$ obtained in Eq. (\ref{rescaled2}). As discussed at the
end of the previous subsection, a meaningful first approximation is to neglect the 
double product prefactor in Eq. (\ref{rescaled2}), while retaining the ordering condition (hence 
accounting for particle crossing). 
The additional effect of this prefactor will 
be discussed in the following section. If we do so we obtain 
% and rewrite it as
% \bea  \label{rescaled3} 
% P(\vec x,t) \simeq \frac{1}{(4 \pi t)^{N/2}} e^{- \frac{t}{4} \sum_{j=1}^N (z_j - c (2 j - N - 1))^2  +\sum_{a<b} \ln(\frac{z_b-z_a}{z_b-z_a+2c})} \;.
%  \eea 
% For fixed $N$ but large $t$, clearly the first term dominates over the second term in the exponent in Eq. (\ref{rescaled3}). Hence, to leading order for large $t$ and any $N$, one can write  
\bea\label{rescaled4}
P(\vec x,t) \propto e^{- \frac{t}{4} \sum_{j=1}^N (z_j - c (2 j - N - 1))^2} \quad, \quad {\rm for} \quad z_1<z_2<\cdots< z_N\;.
\eea
}
% {\blue We emphasize that the neglect of the double product prefactor in Eq. (\ref{rescaled2}) is exact at large $t$ and fixed $N$ only
% if one scales $x_j/\sqrt{t}=O(1)$ (equivalently $z_j \sim 1/\sqrt{t}$).}

We are now ready to compare Eqs. (\ref{bolt_ord}) and (\ref{rescaled4}). We see that the two probability distributions are 
formally equivalent provided we identify 
\be \label{transformation}
z_i = \sqrt{ \frac{2}{t}} N y_i \quad , \quad \alpha = \frac{c \sqrt{t}}{2 \sqrt{2}}  \;,
\ee 
{which also gives $x_i= z_i t = \sqrt{2 t} N y_i$.  
Since our original large time formula (\ref{rescaled2}) was obtained for $c^2 t \gg 1$
we see that the predictions from the equilibrium problem can be translated to the dynamics problem a priori only for $\alpha \gg 1$. 
However, it is interesting
to present and use below some of the known results for the equilibrium problem at 
arbitrary $\alpha$ (with the idea that they may capture some of the effects of particle crossing for $c^2 t=O(1)$).}

{The above considerations, and the correspondence \eqref{transformation}, hold for any $N$. Let us now consider
the case where $N \gg 1$. 
%Note that we are considering here the limit where $t$ is taken large first. In a second stage we now study $N \gg 1$. 
In that case, 
%have assumed that {\blue $c^2 t \gg 1$} for the dynamics problem. 
% according to Eq. (\ref{transformation}), the dynamics maps onto the equilibrium jellium model in the strongly interacting regime $\alpha \gg 1$. Since the density of $y_i$'s is supported over $[-2 \alpha, + 2 \alpha]$ --see Eq. (\ref{rho_lim})--, it is clear that $y_i's$ are of
% order $O(\alpha) \sim O(\sqrt{t})$ for large $t$. Hence, from Eq. (\ref{transformation}), $z_i = O(N)$ for large $N$. Therefore, 
from the average density in the equilibrium problem %in terms of the $y_i$'s variables 
in Eq. (\ref{rho_lim}), and using \eqref{transformation},} we can make a prediction for the density of the $z_i$'s variables (defined similarly to \eqref{def_rho}) in the
dynamics problem, namely
\bea \label{rho_z}
\rho(z)  \approx \frac{1}{2 c\,N} \mathbb{I}_{[-c\,N\,, +c\, N]}(z) \;.
\eea
{The prediction for the density in the original coordinates of the particles, $x_i = t\, z_i$, thus takes the form} %Hence, the density in the $x$-variables, for large $t$ reads
\bea \label{rho_x}
\rho(x,t) \approx \frac{1}{2c\,N\,t}  \mathbb{I}_{[-c\,N\,t, +c\, N \,t]}(x) \;.
\eea 
% \begin{figure}[t]
% \centering
% \includegraphics[width=0.6\linewidth]{Plots/Density_profile1.pdf}
% \caption{Average density of $N=1000$ particles at $t=1000$, $c=0.1$ and $T=1$. The averaging is done over $10^3$ realizations of the noise.
% {\red P: need to comment on the step. What does the point in the middle of the step mean? 
% Also maybe mention that in the present regime the scale $T/(c N)$ is vanishing (does not really exist) or this
% should be discussed later?}} \label{fig:G1}
% \end{figure}
This describes the dynamics of a gas whose two edges move ballistically with constant speed $c\, N$, describing two light-cones that bound
the trajectories of the gas particles [see Fig. \ref{Plot_traj} c) in the middle panel]. {The prediction 
\eqref{rho_x} is in agreement with the density computed numerically and represented in the bottom panel of Fig. \ref{Plot_traj} c).
It corresponds to the regime III discussed there.
}

{Let us now recall for completeness some exact results for various observables that were derived recently for the jellium model for any $\alpha$,
and later consider the large $\alpha$ limit where it leads to predictions for the ranked diffusion at large time.
Consider now the gap between two consecutive particles both in the bulk and at the edges.}
% Given this one-to-one correspondence in Eq. (\ref{transformation}) between the ranked diffusion problem at finite but large $t$ and the equilibrium jellium model in the strongly interacting regime, one can use several exact results for various observables that were derived recently for the jellium model and make precise predictions for the ranked diffusion problem. For this correspondence to work, we need to consider the results for the jellium model in the large $\alpha$ limit. Below, we discuss a precise example of one such observable, namely 
% the gap between two consecutive particles both in the bulk and at the edges. 
Consider %first 
the jellium model in Eq. (\ref{bolt_ord}) and let $g_i=y_{i+1}-y_{i}$ denote the spacing between the $i$-th and $(i+1)$-th particle of the jellium gas at equilibrium. First, we consider the mid-gap, i.e., setting $i=N/2$ (this corresponds to the typical gap in the bulk). In this case, the distribution of the mid-gap, in the large $N$ but fixed $\alpha$ limit, takes the scaling form \cite{Flack22}
\begin{align}\label{bulk_gap}
    \mathcal{P}_{\text{mid-gap}}(g,N) & \sim NH_{\alpha}(gN),\nonumber \\
    H_{\alpha}(z) &= \theta(z)[A(\alpha)]^2\int_{-\infty}^{\infty}dy e^{-\frac{1}{2}[(y+z-4\alpha)^2+y^2]}F_{\alpha}(y+4\alpha)F_{\alpha}(-y-z+8\alpha)\;,
\end{align}
where the function $F_\alpha(x)$ satisfies the non-local differential equation
\bea\label{eq:Falpha}
\frac{dF_\alpha(x)}{dx} = A(\alpha) F_\alpha(x+4\alpha) e^{-\frac{x^2}{2}} \;,
\eea 
with the boundary conditions $F_\alpha(x\to +\infty)=1$ and $F_\alpha(x \to -\infty) = 0$. This equation can be thought of as an eigenvalue 
equation, with $A(\alpha)$ as the unique eigenvalue for which there exists a solution that satisfies both boundary conditions. In particular, for $\alpha$ large, it behaves as $A(\alpha \rightarrow\infty) \sim 1/\sqrt{2\pi}$ \cite{Baxter,SatyaJellium1}. This function $F_\alpha(x)$ often appears in the context of 1dOCP \cite{SatyaJellium1,SatyaJellium2,Flack22} (see also \cite{Baxter}) and it has the following asymptotic behaviors \cite{SatyaJellium1,SatyaJellium2}
\begin{align}
    F_{\alpha}(x) &\sim 1- e^{-x^2/2+ o(x^2)} \quad \quad \; \text{ for } x\rightarrow\infty\;,\label{largex}\\
    F_{\alpha}(x) &\sim e^{-|x|^3/(24 \alpha)+o(x^3)} \quad \quad \text{ for } x \rightarrow -\infty \;. \label{smallx}
\end{align}
{Let us now focus on 
%In order to draw conclusion about the statistics of the gap in the bulk in the dynamics problem, we need to first focus on 
the distribution in Eq. (\ref{bulk_gap}) in the large $\alpha$ limit.} In this limit
we can approximate the integral over $y$ in Eq. \eqref{bulk_gap} by a saddle-point method. The minimum of the argument in the exponential function occurs at $y^*=(4\alpha - z)/2$. At this value of $y$, the $F_\alpha$-functions in the integrand in Eq. (\ref{bulk_gap}) read $[F_\alpha(6\alpha - z/2)]^2$. Since $\alpha$ is large, this factor essentially contributes unity, using Eq. (\ref{largex}), as long as $z<12 \alpha$. We will see in the following that indeed this is true in the range of $z$ where the gap distribution has a peak. 
Therefore the saddle point analysis gives, up to a multiplicative prefactor,
\be \label{H_gauss}
\lim_{\alpha \rightarrow \infty} H_{\alpha}(z) \sim e^{-\frac{1}{4}(z-4\alpha)^2} \;,
\ee
%since the contributions from functions $F(x)$ are approximately one in this limit.
Hence, the distribution of the mid-gap, in the limit of large $\alpha$, approaches a Gaussian distribution 
\be 
\mathcal{P}_{\text{mid-gap}}(g)\sim e^{-\frac{N^2}{4}(g-4\alpha/N)^2} \label{mid-gap-gauss} \;,
\ee
with mean at $g = 4\alpha/N$ and variance $2/N^2$. Thus, in this limit of large $\alpha$ (and large $N$), one can express the random variable $g_{\rm mid-gap}$ (i.e., the mid-gap) as
\bea \label{g_Gauss}
g_{\rm mid-gap} \approx \frac{4 \alpha}{N} + \frac{\sqrt{2}}{N} {\cal N}(0,1) \;,
\eea
where ${\cal N}(0,1)$ is a standard normal variable with zero mean and unit variance. We can now use this result and the correspondence in Eq. (\ref{transformation}) to predict the distribution of the mid-gap $g^{\rm RD}_{\rm mid-gap}$ in the dynamics problem of ranked diffusion (the superscript `RD' refers to ranked diffusion). We get
\be
g^{\rm RD}_{\rm mid-gap} = z_{N/2+1} - z_{N/2} \approx \sqrt{\frac{2}{t}} N \left[ \frac{4 \alpha}{N} + \frac{\sqrt{2}}{N} {\cal N}(0,1) \right] 
=2 c + \frac{2}{\sqrt{t}}{\cal N}(0,1)  \;. \label{g_RD}
\ee
Going back to the original $x_i = t\, z_i$ coordinates, we finally get the mid-gap distribution as
\bea \label{GRD}
G^{\rm RD}_{\rm mid-gap} = x_{N/2+1}(t) - x_{N/2}(t) = t g^{\rm RD}_{\rm mid-gap} \approx  2 c\,t + 2\,\sqrt{t}{\,\cal N}(0,1) \;.
\eea 
{The same result holds for all the gaps inside the bulk} of the jellium model \cite{Flack22}, and hence equivalently for the ranked diffusion model.
\begin{figure}[t]
\centering
\includegraphics[width=0.6\linewidth]{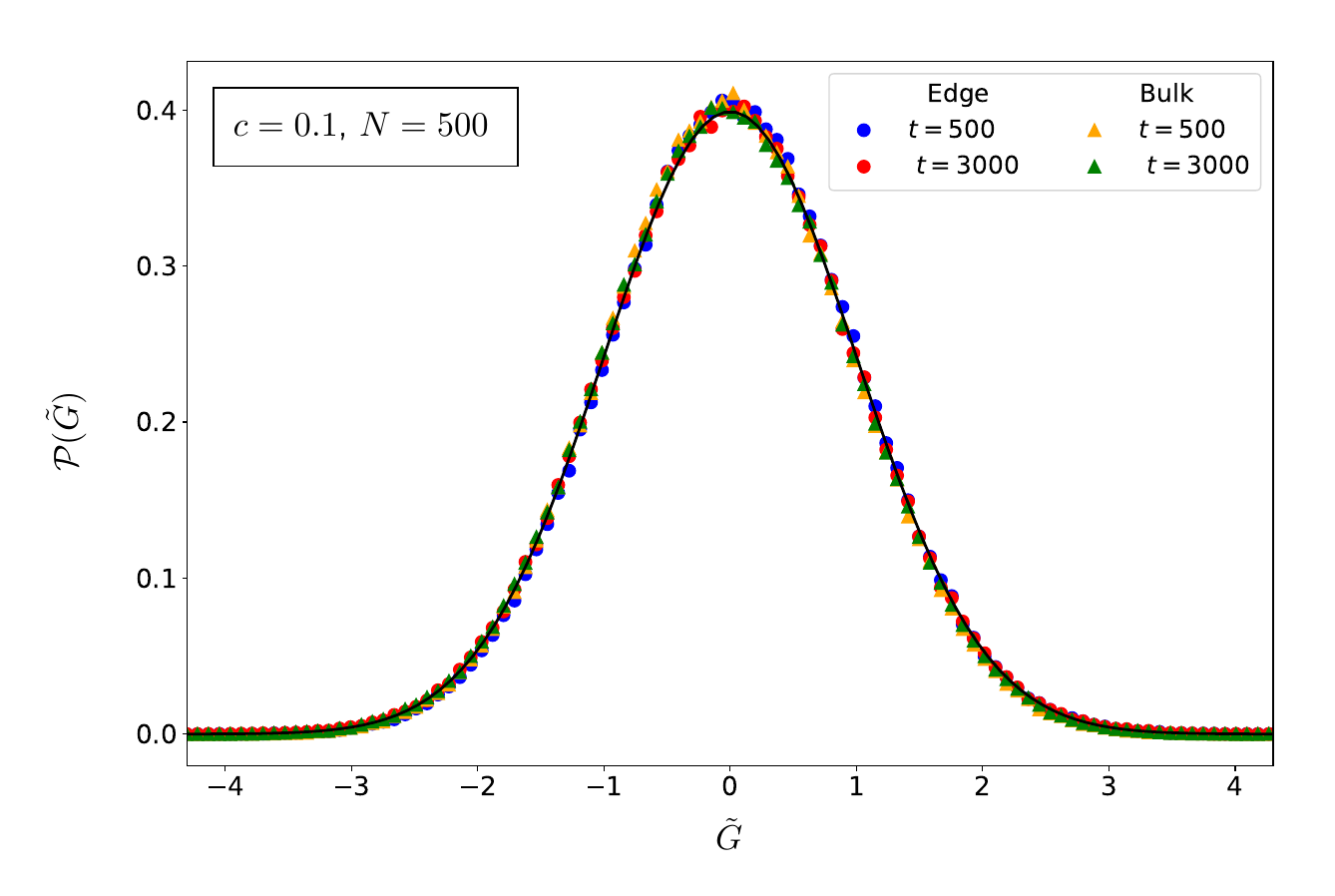}
\caption{Plot of the distribution of the centered and scaled gap 
$\tilde G = \frac{G^{RD}-2 c t}{2 \sqrt{t}}$ for times $t=500$ and $t=3000$, from the numerical simulation of the Langevin equation \eqref{langevin1}
both for the mid-gap $G^{RD} \equiv G^{RD}_{\rm mid-gap}$ and for the edge gap $G^{RD} \equiv G^{RD}_{\rm edge-gap}$. Here 
$N=500$, $c=0.1$ (note that $c^2 t$ is the dimensionless time). It is compared
with the normal Gaussian distribution $e^{-x^2/2}/\sqrt{2 \pi}$ as predicted in \eqref{GRD} and \eqref{GRD_edge} (black color). 
 %Numerical data is calculated for the particles from the edge and particles from the bulk of the system. 
 %The measurements for different times ($t\in \{500, 1000, 1500, 2000\}$ collapse on the same curve $1/\sqrt{2 \pi} e^{-x^2/2}$. 
 %In the simulation we set $N=500$ and $\gamma=N c=50$.
 }\label{fig:G1}
\end{figure}
%While the calculation above is presented exactly for the mid-gap, it actually 
However, the behavior of the gap distribution changes as one approaches the edges of the jellium model, 
i.e., 
%In fact, it is known that the gap at the edge of the jellium model, 
$g_{\rm edge-gap} = y_{N} - y_{N-1}$, has the following distribution in the large $N$ limit \cite{SatyaJellium2}
\begin{align}\label{edge_gap}
    \mathcal{P}_{\text{edge-gap}}(g) & \sim N\, h_{\alpha}(gN),\\
    h_{\alpha}(z) &= \theta(z)[A(\alpha)]^2\int_{-\infty}^{\infty}dy e^{-\frac{1}{2}[(y+z-4\alpha)^2+y^2]}F_{\alpha}(y+4\alpha),
    \label{eq:gap}
\end{align}
where $F_\alpha(x)$ is the same function as defined in Eq. (\ref{eq:Falpha}). Once again, to make correspondence with the ranked diffusion problem, we need to consider the limit $\alpha$ large. In that limit, one can again approximate this integral (\ref{eq:gap}) by the saddle-point method. Following exactly the same argument as in the mid-gap case, one gets 
\be 
\lim_{\alpha \rightarrow \infty} h_{\alpha}(z) \sim \frac{1}{\sqrt{4 \pi}} e^{-\frac{1}{4}{(z-4 \alpha})^2} \;.
\ee
Hence, using (\ref{edge_gap}), we again get a Gaussian distribution for the edge gap in the jellium model in the large $\alpha$ limit
\be 
\mathcal{P}_{\text{edge-gap}}(g)\sim e^{-\frac{N^2}{4}(g-4\alpha/N)^2} \label{mid-gap-gauss} \;.
\ee
Thus, in the large $\alpha$ limit, the edge and bulk gaps in the jellium model behave in the same way, namely 
\bea \label{gedge_Gauss}
g_{\rm edge-gap} \approx \frac{4 \alpha}{N} + \frac{\sqrt{2}}{N} {\cal N}(0,1) \;.
\eea
Correspondingly, the edge-gap and in the mid-gap behave in a same way, namely 
\bea \label{GRD_edge}
G^{\rm RD}_{\rm edge-gap} = x_{N}(t) - x_{N-1}(t) \approx  2 c\,t + 2\,\sqrt{t}{\,\cal N}(0,1) \;.
\eea 
Although this result, together with \eqref{GRD}, was obtained here at large $N$, we note by comparing to Eq. \eqref{var_N2} that it already holds for $N=2$ for the first two cumulants, which thus appears to independent of $N$. 

{In summary, the predictions \eqref{GRD} and \eqref{GRD_edge} from the analogy with the equilibrium are in agreement with
the discussion at the end of the previous subsection, i.e., that on the scale $\delta x_i \sim \sqrt{t}$ the
fluctuations are independent Gaussian, as indeed recovered in the large $\alpha$ limit of the jellium. 
In Fig. \ref{fig:G1}, we verify by Monte-Carlo simulations the two analytical predictions for the gaps in the ranked diffusion model respectively in Eqs. (\ref{GRD}) and (\ref{GRD_edge}). We see that the agreement is indeed excellent if the gaps are scaled by $\sqrt{t}$. 

However, this is not the end of the story for the dynamics problem, and in the next subsection
we will compute the higher cumulants of the particle positions $x_i$ at large time, which exhibit
deviations from this leading Gaussian behavior.}

%This means that no matter the position of the gap (bulk or edge) we expect Gaussian shaped typical fluctuations centered around zero $\tilde{z}=0$. 
%By correct mapping of the observables we can compare the distributions of the gap in 1dOCP to the one in ranked diffusion.
%From the relations Eq. \eqref{eq:PDF_plasma} and Eq. \eqref{eq:PDF_RD} we can see extract the average gap for both cases
%\begin{align}
%\langle N g^{1dOCP}\rangle=\langle N (x_{i+1}-x_i)\rangle = 4\alpha,\\
%\langle \sqrt{\frac{t}{2}}g^{RD}\rangle=\langle \sqrt{\frac{t}{2}}(z_{i+1}-z_i)\rangle =\frac{2c\sqrt{t}}{\sqrt{2}}  = 4\alpha,
%\end{align}
%where we applied the transformation from Eq. \eqref{transformation} in the last line. 
%Consequently, we compared the distributions of $\tilde{z}^{1dOCP}=Ng^{1dOCP}-4\alpha$ and $G^{RD}=\sqrt{\frac{t}{2}}g^{RD}-\sqrt{2t}c$. 

{ 
\subsection{More accurate treatment of the large time limit: higher cumulants} 
\label{subsec:cumulants} 

In this section we go back to the complete asymptotic form for the joint PDF at large time \eqref{rescaled},\eqref{rescaled2}
and we obtain all the cumulants of the particle positions $x_i$ to $O(1)$ accuracy.

\subsubsection{The case $N=2$} 

Let us start with $N=2$ for simplicity. Consider the large time asymptotic formula \eqref{N_2_z}. 
Let us recall that the original variable is $y=x_2-x_1=z t$. We can consider the sector $y>0$: 
indeed we will use a saddle point method, and there will be one saddle point inside each sector and 
the result will not depend on the sector (see below). We want to evaluate the cumulant generating function 
\be 
\langle e^{\lambda y} \rangle = \langle e^{t \lambda z} \rangle 
 \simeq \int_0^{+\infty} dz  e^{ - \frac{t}{4} (z-2 c)^2 + \log(\frac{z}{z+2 c}) + f(t) + t \lambda z}  
\ee 
where we have inserted \eqref{N_2_z} and the normalization factor $f(t) = - \ln(2 \sqrt{2 \pi t})$ is immaterial for the following.
At large $t$ there is a unique saddle point at $z = z^* = 2 c + 2 \lambda$. We require that $2 c + 2 \lambda>0$
(in fact for the cumulants we only need $\lambda$ in a neighborhood of $\lambda=0$). From the
saddle point method we thus obtain
\be \label{gener2} 
\langle e^{\lambda y} \rangle \simeq \exp \left(  2 t \lambda + 2 \lambda^2 t + \log\left(\frac{2 c +  2 \lambda}{2 c  + \lambda } \right) \right)
\ee 
where we have fixed the normalization so that the r.h.s. is equal to unity at $\lambda=0$. Using
that $\log \langle e^{\lambda y} \rangle = \sum_{k\geq 1} \frac{\lambda^k}{k!} \langle y^k \rangle_c$
and expanding in $\lambda$ we obtain all the cumulants. This reproduces the results in Eqs. 
\eqref{meanN2}-\eqref{cum4N2} up to and including $O(1)$ terms at large time, and gives the more general formula
for $k \geq 2$
\be \label{generalcum} 
\langle y^k \rangle_c = \frac{1}{c^k}\, (-1)^{k-1} (k-1)! (2^k-1) + o (1) 
\ee 
Several remarks are in order. First, since \eqref{N_2_z} depends only on $|z|$, instead of choosing the sector $y>0$ we 
could have done the exact same calculation replacing $y$ by $|y|$, and $z$ by $|z|$. The saddle point 
is then at $|z| = 2 c + 2 \lambda$, so there are in fact two identical saddle points for each sign of $y$. 
Hence to the same accuracy $\langle (|y|-2 c t)^k \rangle_c = \langle y^k \rangle_c$ given by 
\eqref{generalcum}. Next we see that \eqref{gener2} holds for any $\lambda > - c$, but fails when $\lambda \leq -c$, 
since for $\lambda=-c$ the saddle point reaches $z=0$. 
The average $\langle e^{\lambda y} \rangle$ is then dominated by the vicinity of $z=0$, i.e., by events which involve particle
crossings, and the two sectors cannot be neatly separated. Finally, we know from 
the exact results \eqref{meanN2}-\eqref{cum4N2} that the $o(1)$ corrections in 
\eqref{generalcum} should be exponentially small $O(e^{-c^2 t/2})$ at large time. But the above saddle-point
method, if pushed to next order, will lead power law in time corrections $O(1/t^k)$. This apparent paradox is resolved
in the Appendix, where it is shown that \eqref{gener2} has indeed only exponentially small corrections in time.
The reason for that is that \eqref{N_2_z} itself comes from a first saddle point method, and both
saddle points should be considered simultaneously. In the Appendix we identify these exponentially small corrections to 
\eqref{gener2} to come precisely from particle crossing, which are exponentially rare for $c^2 t \gg 1$.

\subsubsection{The case of arbitrary $N$}

Let us now turn to arbitrary $N$ and consider the sector $x_1 < x_2 < \dots <x_N$, recalling that we
denote $x_j= t z_j$. Let us compute the following generating function at large time, inserting the asymptotic form \eqref{rescaled2},
\be 
\langle e^{\sum_j \lambda_j x_j} \rangle = \langle e^{t \sum_j \lambda_j z_j} \rangle \simeq
\int_{z_1<z_2<\dots<z_N}  dz_1 \dots dz_N e^{- \frac{t}{4} \sum_{j=1}^N (z_j - c (2 j - N - 1))^2  + t \sum_{j=1}^N \lambda_j z_j
+ \sum_{1 \leq a < b \leq N} \log \frac{z_b-z_a}{z_b - z_a + 2 c} + f(t) } 
\ee
where again $f(t)$ is an unimportant normalization. In this sector there is a unique saddle point at large $t$ given by
\be 
z_j = z_j^* = c (2 j - N - 1) + 2 \lambda_j \quad , \quad j=1,\dots,N
\ee 
We will assume that the $z_j^*$ are in the sector considered, i.e., that all $c (2 j - N - 1) + 2 \lambda_j >0$
for $j=1,\dots,N$, which is certainly the case when the $\lambda_j$'s are all in a neighborhood of zero. 
Then the saddle point method gives
\be \label{generN} 
\langle e^{\sum_j \lambda_j x_j} \rangle \simeq N! \exp \left(  \sum_{j=1}^N c (2 j - N - 1) t \lambda_j + t \sum_{j=1}^N \lambda_j^2 
+ \sum_{1 \leq a < b \leq N} \log( \frac{c( b-a ) + \lambda_b-\lambda_a}{c + (b-a) c +  \lambda_b-\lambda_a} ) \right)
\ee 
where we used $\sum_{1 \leq a < b \leq N} \frac{b-a}{1+b-a}=1/N!$ to normalize the formula. 

Upon expanding the logarithm of \eqref{generN} in the parameters $\lambda_j$ we can now compute all the joint cumulants of
the deviations from the perfect crystal, defined as $\delta x_j = x_j - c (2 j - N - 1) t$. 
First we note that the double sum in \eqref{generN} involves only pairs of distinct variables $\lambda_j$.
Hence the cumulants involving more than two particles are zero, e.g.
\be 
\langle \delta x_i \delta x_j \delta x_k \rangle_c = 0 \quad , \quad \text{for} \quad i<j<k 
\ee 
To compute the only non-zero cumulants (i.e., involving only one or two particles) we first define the
function 
\be 
f_k(x) = \log \left( \frac{c k+x}{c +c k+x} \right)\;, 
\ee 
which has derivatives
\be 
c^k f_k^{(n)}(0) = (-1)^{n-1} (n-1)! \left( \frac{1}{k^n } - \frac{1}{(k+1)^n } \right) 
\ee 
From the logarithm of \eqref{generN} we obtain the single particle cumulants as
\bea 
&& \langle (\delta x_i)^n \rangle_c = \sum_{i < j \leq N} \partial_{\lambda_i}^n f_{j-i}(\lambda_j-\lambda_i)|_{\lambda_i=\lambda_j=0}
+ \sum_{1\leq j<i} \partial_{\lambda_i}^n f_{i-j}(\lambda_i-\lambda_j)|_{\lambda_i=\lambda_j=0} + 2  t \delta_{n,2}  \\
&& = \sum_{i < j \leq N} (-1)^n f_{j-i}^{(n)}(0) + \sum_{1\leq j<i} f_{i-j}^{(n)}(0) + 2 t \delta_{n,2} 
\eea
This leads to, for $1 \leq i \leq N$ and $n \geq 2$
\be
 \langle (\delta x_i)^n \rangle_c = \frac{1}{c^n} (n-1)! \left(  \frac{1}{(N-i+1)^n}-1 + (-1)^{n-1}(1- \frac{1}{i^n}) \right) + 2 t \delta_{n,2}
\ee 
We see that at large $N$, for $i=O(1)$ i.e., at the left edge of the gas, one has
\be
 \lim_{N \to +\infty, i=O(1)} \langle (\delta x_i)^n \rangle_c = \frac{1}{c^n} (n-1)! \left( -2 \delta_{n,\rm even} + \frac{(-1)^n}{i^n} \right) + 2  t \delta_{n,2}
%\right)
\ee 
We see that in the bulk of the gas, i.e., for $i \to +\infty$, the odd cumulants decay to zero, while the even cumulants $n \geq 4$ decay from
$-\frac{3}{c^n} (n-1)!$ for $i=1$, to a finite limit
$-\frac{2}{c^n} (n-1)!$, which is uniform over the bulk of the gas. Hence the fluctuations are
slightly larger near the edge, but remain finite in the bulk. On the right edge the 
result is similar, with the term $\frac{(-1)^n}{i^n}$ replaced by $\frac{1}{(N-i+1)^n}$.
\\

Next, we compute the cumulants involving two particles (for $j > i$, $m,n \geq 1$ and $N \geq 2$). One finds
\bea 
&& \langle \delta x_i^n \delta x_j^m \rangle_c = \partial_{\lambda_i}^n \partial_{\lambda_j}^m f_{j-i}(\lambda_j-\lambda_i)|_{\lambda_i=\lambda_j=0}
= (-1)^n f_{j-i}^{(n+m)}(0) \nn \\
&& = \frac{1}{c^n} 
(-1)^{m-1} (n+m-1)! \left( \frac{1}{(j-i)^{n+m} } - \frac{1}{(j-i+1)^{n+m} } \right)  \label{cumul1}
\eea 
We note that these correlations are independent of $N$ and decay quickly to zero as a function of the distance between the two particles. 

These formulae are exact for all $N$. For concreteness let us display some of the predictions for $N=2,3$. For simplicity we set here $c=1$.
For $N=2$ this gives 
\bea 
&& \langle (\delta x_1)^2 \rangle_c = \langle (\delta x_2)^2 \rangle_c = 2 t - \frac{3}{4} 
\quad , \quad \langle \delta x_1 \delta x_2 \rangle_c = \frac{3}{4} \\
&& \langle (\delta x_1)^3 \rangle_c =  - \langle (\delta x_2)^3 \rangle_c = -  \frac{7}{4} 
\quad , \quad \langle \delta x_1 (\delta x_2)^2 \rangle_c = - \langle (\delta x_1)^2 \delta x_2 \rangle_c =  - \frac{7}{4} 
\eea 
For $N=3$ we obtain
\bea 
&& \langle \delta x_1 \rangle = -2/3 \quad , \quad \langle \delta x_2 \rangle = 0 \quad , \quad \langle \delta x_3 \rangle_c = 2/3 \\
&& \langle \delta x_1^2 \rangle_c = \langle \delta x_3^2 \rangle_c  = 2 t - \frac{8}{9} \quad , \quad \langle \delta x_2^2 \rangle_c = 2 t - \frac{3}{2}
\quad , \quad \langle \delta x_1 \delta x_2 \rangle_c = \langle \delta x_2 \delta x_3 \rangle_c = \frac{3}{4}
\quad , \quad \langle \delta x_1 \delta x_3 \rangle_c = \frac{5}{36}\\
&& \langle \delta x_1^3 \rangle_c =  -  \langle \delta x_3^3 \rangle_c  = - \frac{52}{27} \quad , \quad \langle \delta x_2^3 \rangle_c = 0
\quad , \quad \langle \delta x_1 \delta x_2^2 \rangle_c = - \langle \delta x_1^2 \delta x_2 \rangle_c = - \frac{7}{4} 
\\
&& \langle \delta x_1^2 \delta x_3 \rangle_c = - \langle \delta x_1 \delta x_3^2 \rangle_c =  \frac{19}{108} \quad , \quad  \langle \delta x_2 \delta x_3^2 \rangle_c = - \langle \delta x_2^2 \delta x_3 \rangle_c = - \frac{7}{4}  
\quad , \quad \langle \delta x_1 \delta x_2 \delta x_3 \rangle_c = 0 \;.
\eea

Next one can use these formula to compute the cumulants of the relative distance between any two particles. Since the
cumulants of $x_j$ and of $\delta x_j$ are by definition identical, we will simply use the particle positions $x_j$ here.
For $1 \leq i < j \leq N$ one finds
%{\red P: new formula, Greg can you check it? }
\bea  
&& \langle (x_{j}-x_i)^n \rangle_c = \frac{(n-1)!}{c^n}  \bigg(i^{-n}+(-1)^n
   \left(2^n-2\right)
   \left((-i+j+1)^{-n}-(j-i)^{-n}\right) \\
   && +(-1)^n (-i+N+1)^{-n}+(-1)^n
   j^{-n}+(-j+N+1)^{-n}-2
   (-1)^n-2\bigg) + 4 t \delta_{n,2} + 2 t {c \, (j-i)}\delta_{n,1} \;,
   \eea  
which for $j=i+1$ gives the cumulants of the gaps, for which one finds for $1 \leq i \leq N-1$
%{\red P: Greg I simplified a bit can you check? old formula is commented} 
%old formula commented 
% \bea
% && \langle (x_{i+1}-x_i)^n \rangle_c =
% (n-1)! 
%    \bigg(i^{-n}+(N-i)^{-n}+(-1)^n
%    (-i+N+1)^{-n}+(-1)^n
%    (i+1)^{-n} \\
%    && +(-1)^{n+1}
%    2^{1-n}+(-1)^n+(-1)^{n+1} 2^n-2\bigg) + 4 t \delta_{n,2} + 2 t \delta_{n,1} 
% \eea 
\bea   \label{gapscum} 
&& \langle (x_{i+1}-x_i)^n \rangle_c = \frac{(n-1)!}{c^n} 
   \bigg(i^{-n}+(N-i)^{-n}+(-1)^n
   (-i+N+1)^{-n}+(-1)^n (i+1)^{-n} \\
   && {+} (-1)^n \left(-2^{1-n}-2^n+1\right)-2 \bigg) + 4 t \delta_{n,2} + 2 c\,t \delta_{n,1} \;. \nn
\eea  
Again, in the large $N$ limit one finds some distinct behavior at the two edges,
and that the cumulants of the gap reach a finite limit inside the bulk given by
the second line of \eqref{gapscum}.

Similarly one finds that the cumulants of the total size of the gas (i.e., its span) are given by
%{\red P: Greg I simplified a bit can you check? old formula is commented} 
% old formula commented
%\bea 
%&& \langle (x_{N}-x_1)^n \rangle_c = 
%(-1)^n N^{-n} \left(2^n-\left(2
%    (N-1)^n+2^n-2\right) (N-1)^{-n}
%    N^n\right) \Gamma (n) + 4 t \delta_{n,2} + 2 N t \delta_{n,1}  
% \eea 
\bea \label{span}
\langle (x_{N}-x_1)^n \rangle_c =
(-1)^n \frac{(n-1)!}{c^n}  \left(2^n
   N^{-n}{-} \left(2^n-2\right)
   (N-1)^{-n}{+}2\right)  + 4 t \delta_{n,2} + 2 {c\,(N-1)} t \delta_{n,1}  \;. 
\eea

For large $N$ the $n$-th cumulant ($n >1$) of the total size of the gas converge quickly to some finite limit
\bea
\langle (x_{N}-x_1)^n \rangle_c \simeq 2 (-1)^{n-1} \frac{(n-1)!}{c^n}  (1 - \frac{1}{N^n} + \dots) + 4 t \delta_{n,2} + 2 c (N-1) \, t \delta_{n,1}  \;,
\eea
which, at large $N$, is equivalent to the sum of independent fluctuations, $\langle (\delta x_N)^n \rangle_c + (-1)^n \langle (\delta x_1)^n \rangle_c$,
since the mutual fluctuations decay at large distance as pointed out above.
It is interesting that there are non Gaussian persistent correlations at large time, 
on the scale of the gas, even at large $N$. 

To summarize we have obtained a complete quantitative picture of the $O(1)$ fluctuations at large time in the
expanding crystal which goes beyond the independent Gaussian $O(\sqrt{t})$ fluctuations discussed in the
previous subsections. These result are valid for $c^2 t \gg 1$ and any $N$. From the considerations
for $N=2$ (see Appendix \ref{app:twoparticles}), we can surmise that the corrections to these cumulants, and to \eqref{generN}, are
exponentially small at large time and related to particle crossing (hence also in part to the
finite $\alpha$ physics of the equilibrium jellium).
}

%{\red P: Maybe one can also do linear statistics of the type $t \sum_i f(z_i)$}

\section{Approach via the Burgers equation} \label{sec:burgers}

{Until now we have focused on taking the large time limit first, with a fixed number of particles $N$. In this section we first recall the exact hydrodynamic equation which describes the evolution of the density. Using this equation we then study the limit of large $N$ first, at arbitrary fixed time $t$. {This provides
a description of the dense regimes I and II discussed in the introduction.} Finally, we compare the predictions with numerical simulations.}

\subsection{Burgers equation and large $N$ limit} 

Let us first recall the general approach developed in \cite{PLDRankedDiffusion}. 
Let us consider again the Langevin equation \eqref{langevin1} for $N$ particles.
One defines respectively the density field $\rho(x,t)$ and the rank field 
$r(x,t)$ as
\be \label{defrank} 
\rho(x,t) = \frac{1}{N} \sum_i \delta(x-x_i(t)) = \partial_x r(x,t) \quad , \quad r(x,t) = \int^x_{-\infty} dx' \rho(x',t) - \frac{1}{2} \;.
\ee
It is convenient to choose the rank field $r(x,t)$ increasing monotonically from $-1/2$ at $x=-\infty$ to $+1/2$ at $x=+\infty$.
Then it is shown in \cite{PLDRankedDiffusion}, using the Dean-Kawasaki method \cite{KK93,Dean,Kawa}, that 
the rank field satisfies the stochastic equation
\be \label{eqr} 
 \partial_t r(x,t) = T \partial_x^2 r(x,t)
- 2 N c \, r(x,t) \partial_x r(x,t)  + \frac{1}{\sqrt{N}}  \sqrt{ 2 T \partial_x r(x,t) } \eta(x,t) \;.
\ee 
In the right hand side (RHS) of (\ref{eqr}), the first term originates from diffusion, the second is a convection term where the local velocity is proportional to the local rank, while the third term is the noise, originating from local Brownian dynamics. 
In \cite{PLDRankedDiffusion} the case of an additional external potential $V(x)$ was also considered, 
but here we set it to zero. Hence Eq. \eqref{eqr} is the Burgers equation with a multiplicative noise. Note that the
function $r(x,t)$ is constrained to be increasing in $x$, so that the density remains positive. This equation is
formally exact for arbitrary $N$ (with the possible mathematical caveat that $r(x,t)$ is a discontinuous stochastic function).
Here we recall that we consider $c>0$.

Let us now discuss the large $N$ limit. {As discussed in \cite{PLDRankedDiffusion} there are a priori two natural scalings of $c$ in that limit.} 
In both cases, the noise term is formally subdominant.

\subsection{Detailed solution for $c=\gamma/N$ and comparison with numerics} 
\begin{figure}[t]
    \centering
    \includegraphics[width=\linewidth]{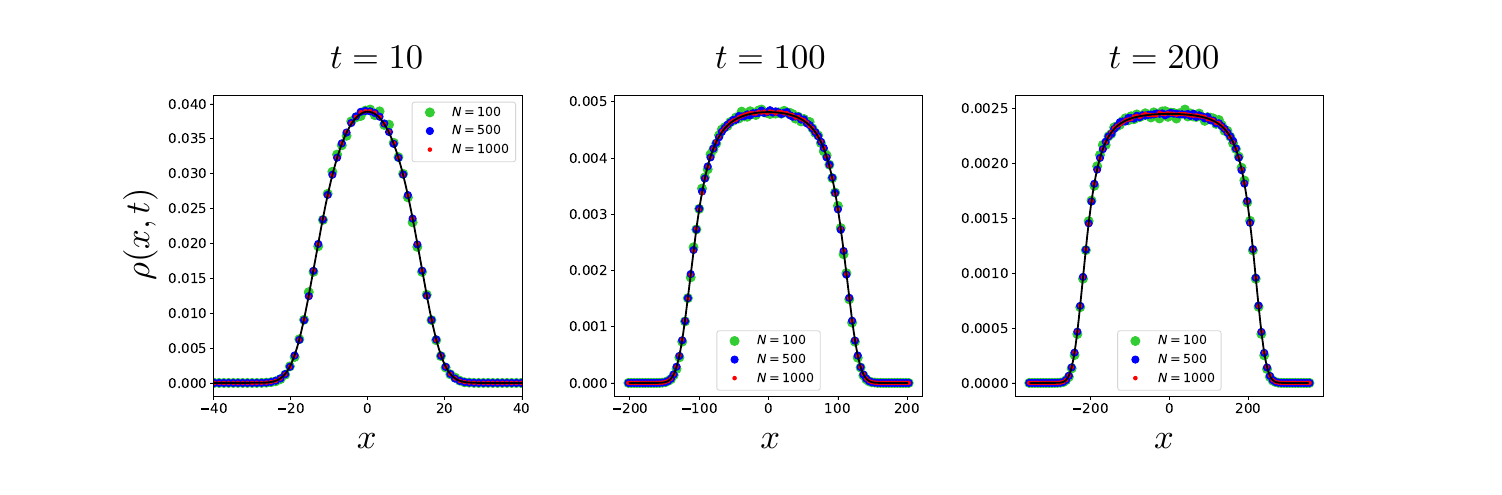}
    \caption{Plot of the density $\rho(x, t)$ evaluated numerically from the Langevin equation, at three different times $t=10, 100$ and $t=200$ for $\gamma=1$ and $T=1$. The initial condition is that all particles are at the origin at time zero. 
For each time we simulated the density profile for three different combinations of $N$ and $c$ with fixed $\gamma=Nc=1$. The averaging is done over $10^4$ realizations of the noise. The numerical data is compared with the analytical prediction with a delta function initial condition from Eq. \eqref{burgers1} (black solid lines).}
    \label{fig:time_densities}
\end{figure}
(i) The first choice is to keep $c$ fixed. In that case, one defines a rescaled time $\tau=N t$. The noise term
and the diffusion term become both $O(1/N)$ and \eqref{eqr}
simply becomes the inviscid Burgers equation, $\partial_\tau r = - 2 c r \partial_x r$. The solution is obtained implicitly by solving for $r \equiv r(x,t)$ equation~\cite{PLDRankedDiffusion}
\be
r = r_0(x - 2 c \, r \tau)   %\quad \Longleftrightarrow \quad r=r(x,t) 
\ee
where $r(x,0)=r_0(x)$ is the initial condition. Since $c>0$ there is a unique solution
with no shocks. An example is the square density 
initial condition, $\ell>0$
\bea \label{solurep0}
\rho(x, t) = \frac{1}{2 (\ell+ c \tau)} \theta(\ell + c \tau-|x|) \quad, \quad \tau = N\,t \;,
\eea 
which shows that the repulsive gas expands linearly in time with sharp edges at $\pm c \tau = \pm c N t$. This result (restoring $\tau = N t$) {matches} with the prediction \eqref{rho_x} which was obtained in the large time limit followed by the large $N$ limit. {Note that here the
convergence to the form \eqref{solurep0} occurs quite fast, on a time scale $\tau=O(1)$ that is $t \sim 1/N$.
Comparing with the discussion in the introduction, we see that this time regime corresponds to regime II where the square density
forms and expands, and in the inviscid Burgers equation the edges are sharp on scale $x=O(1)$.}

(ii) Here we will consider the second and richer choice of scaling, i.e., $c= \gamma/N$ where $\gamma>0$ is fixed.
In that case, only the noise term is subdominant $O(1/\sqrt{N})$ in \eqref{eqr}, 
and one obtains the viscous Burgers equation in the original time variable $t$, namely 
\be \label{eqr2} 
 \partial_t r(x,t) = T \partial_x^2 r(x,t) - 2 \gamma \, r(x,t) \partial_x r(x,t)  \;.
\ee 
As we will see below, the solution of this equation with a square density initial condition (with $\ell>0$) is similar in the bulk to \eqref{solurep0}, since the first term in the RHS of \eqref{eqr2} is essentially zero for a flat density profile. This also leads to an
expansion of the gas which is linear in time with the same speed. The main difference occurs near the edges, the sharp edges of (\ref{eqr}) at $\pm \gamma t = \pm c\,N\,t$ being replaced by a smooth profile with a boundary layer form. For attractive interactions studied in \cite{PLDRankedDiffusion} there
is a stationary state and the boundary layer has a width $\xi = \frac{T}{c N} = \frac{T}{\gamma}$ determined
by comparing the two terms in the r.h.s. of \eqref{eqr2}.
% In some sense this scaling amounts to zoom in on the details of the solution \eqref{solurep0}. 
It turns out that in the case studied here, i.e., repulsive interactions and a delta initial condition for the density,
the gas is always far from stationarity and we show below that the scale which determines the size of the boundary layer 
is the diffusion length scale $\ell_T = \sqrt{2 T t}$ (within the boundary layers the three terms in
\eqref{eqr2} are of the same order). 

Note that the characteristic length and time scales associated with the Burgers equation \eqref{eqr2} 
are $x \sim T/\gamma= T/(N c)$ and $t \sim T/\gamma^2= T/(N^2 c^2) = t_1^*$ (which would allow to eliminate
the $\gamma$ dependence in \eqref{eqr2}). Hence this equation naturally describes the crossover between
the regime I and II.

% are $t =T \tilde t/\gamma^2$
% and $x= T \tilde x/\gamma$, such that in terms of the variables $\tilde x$, $\tilde t$, Eq. \eqref{eqr2} holds with the parameter $\gamma=1$
% and $T=1$. Hence we are looking here at time scales of order $t = O(t_1^*)$ with $t_1^* = T/(c N)^2$. 
% Here the choice $c=\gamma/N$ implies that the shortest time scale defined in the introduction $t_1^*=1/\gamma=O(1)$,
% which means that it is well suited to study the crossover from regime I to regime II. 
% }

% {Finally it is important to note that the noise term is proportional to $\sqrt{\rho(x,t)}$. Hence it can be neglected
% uniformly in $N$ at large $N$ only if the initial density is bounded (for $c>0$ it remains so for all times). This
% point will be discussed again below.} 

In order to compare with the numerics we give here the explicit form of the density and rank field obtained by solving analytically Eq. \eqref{eqr2}
for two cases.

\begin{figure}[t]
\centering
\includegraphics[width=0.6\linewidth]{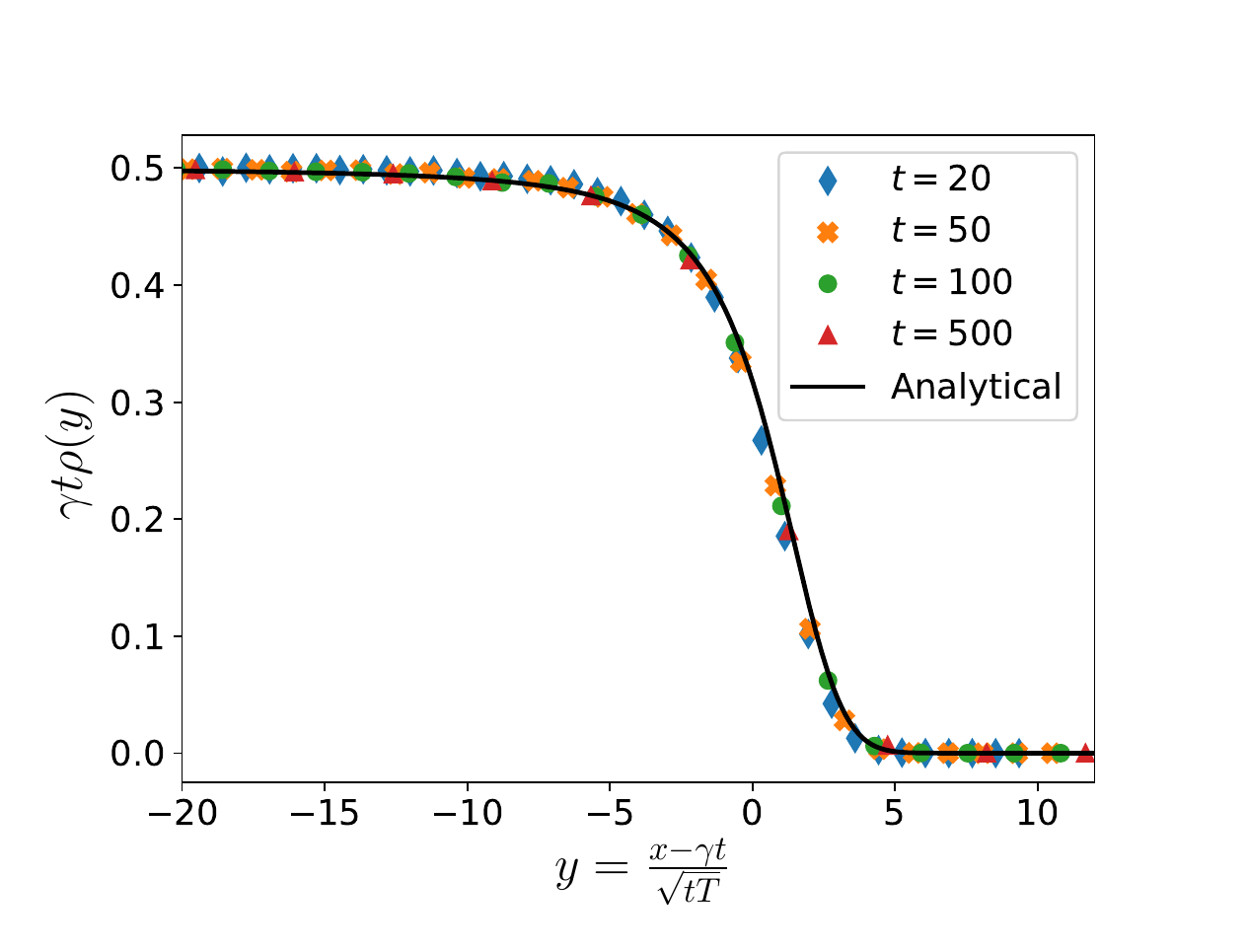}
\caption{Comparison of numerical simulation of the boundary layer at the right edge with the result from Eq. \eqref{BL}. In the simulation $\gamma=cN=7$ with $N=1000$, $c=0.007$ and $T=1$. We chose four different times $t\in \{20, 50, 100, 500\}$, which is inside regime II. The averaging is done over $10000$ realizations of the noise and at $t=0$ all particles are at $x=0$. } \label{fig:BL}
\end{figure}
(i) For the initial condition corresponding to all particles initially at $x=0$, as in \eqref{delta},
which corresponds to $r_0(x)=\frac{1}{2} {\rm sgn}(x)$
one finds (e.g. using Eq. (25) in \cite{PLDRankedDiffusion}) 
\bea \label{burgersdelta} 
&& r(x,t)= - \frac{T}{\gamma} \partial_x \log \left( f(x,t) + f(-x,t) \right) \\
&& f(x,t) = \int_0^{+\infty} \frac{dw}{\sqrt{4 \pi T t}} e^{ - \frac{(w-x)^2}{4 T t} - \frac{\gamma}{T} \frac{w}{2} } 
= \frac{1}{2} e^{\frac{1}{4 T}
   \gamma (\gamma t-2 x)}
   \text{erfc}\left(\frac{\gamma t-x}{2
   \sqrt{T t}}\right) \;.
\eea 
This leads to the time dependent density
\bea \label{burgers1}
\rho(x,t) = \partial_x r(x,t) = 
\frac{e^{-\frac{(x-\gamma  t)^2}{4 t T}}
   \left(\frac{e^{\frac{\gamma  x}{T}}
   \text{erfc}\left(\frac{\gamma  t+x}{2 \sqrt{t}
   \sqrt{T}}\right)}{\sqrt{t}
   \sqrt{T}}+\text{erfc}\left(\frac{\gamma  t-x}{2
   \sqrt{t} \sqrt{T}}\right) \left(\frac{1}{\sqrt{t}
   \sqrt{T}}-\frac{\sqrt{\pi } \gamma 
   e^{\frac{(\gamma  t+x)^2}{4 t T}}
   \text{erfc}\left(\frac{\gamma  t+x}{2 \sqrt{t}
   \sqrt{T}}\right)}{T}\right)\right)}{\sqrt{\pi }
   \left(\text{erfc}\left(\frac{\gamma  t-x}{2
   \sqrt{t} \sqrt{T}}\right)+e^{\frac{\gamma  x}{T}}
   \text{erfc}\left(\frac{\gamma  t+x}{2 \sqrt{t}
   \sqrt{T}}\right)\right)^2} \;.
   \eea 
%{Note that the rounding scale $\xi$ in \eqref{xi} {appears naturally} in this formula.}
\\

(ii) For a box shape initial density of width $2 \ell$, i.e., $\rho(x,0)=\frac{1}{2\ell} \theta(\ell-|x|)$, 
the solution is given by (see Eqs. (171)-(172) in \cite{PLDRankedDiffusion}), where one sets $T=1$
\bea \label{burgersbox} 
&& \rho(x,t) = \partial_x r(x,t) \quad , \quad  r(x,t)= - \frac{1}{\gamma} \partial_x \log \left( f(x,t) + f(-x,t) \right) \\
&& f(x,t)= \frac{e^{-\frac{\gamma x^2}{4 ({\ell}+\gamma
   t)}}}{\sqrt{1+\frac{\gamma t}{{\ell}}}} 
 \text{erf}\left(\frac{{\ell}+\gamma
   t+x}{2 \sqrt{t} \sqrt{1+\frac{\gamma  t}{{\ell}}}}\right)
   +e^{\frac{1}{4}
   \gamma ({\ell}+\gamma t+2 x)}
   \text{erfc}\left(\frac{{\ell}+\gamma t+x}{2
   \sqrt{t}}\right) \;.
\eea 
For $\ell \to 0$ this formula gives back the result \eqref{burgersbox} for the delta initial condition.

The analytical formula \eqref{burgers1} for the density, with a delta initial condition, is
plotted in Fig. \ref{fig:time_densities} for several values of $t$ (solid line). We see
that in the limit of large time $t \gg \frac{1}{\gamma^2}$ the density predicted by the Burgers equation
evolves towards a flat profile for $x \in [-\gamma t, \gamma t]$. In addition, as
can be seen on the figure, there is a boundary layer at each edge. From \eqref{burgers1}
one can derive the precise form of this boundary layer (see Appendix \ref{app:bl})
and one finds that the density near the right edge at $x = \gamma t$ takes the form
\be   \label{BL}
\rho(x,t) = \frac{1}{\gamma t} \hat \rho\left(\frac{x-\gamma t}{\sqrt{T t}}\right) 
\quad , \quad \hat \rho(y) = \frac{e^{-\frac{y^2}{2}}  \left(2+\sqrt{\pi } e^{ \frac{y^2}{4}} y \, \text{erfc}\left(-\frac{y}{2}\right)\right)}{2 \pi \text{erfc}\left(-\frac{y}{2}\right)^2} \;.
\ee   
The characteristic width is thus the diffusion length $\ell_T= \sqrt{T t}$, see \eqref{length}.
As mentioned above, here $t = O(T/\gamma^2)=O(T/(c N)^2)$, hence this width is itself of
order $T/(c N)$. The scaling function has the asymptotic behavior $\hat \rho(y) \simeq \frac{1}{2} - \frac{1}{y^2}$ 
for $y \to -\infty$, which thus matches the density of the plateau $\rho(x,t) \simeq \frac{1}{2 \gamma t}= \frac{1}{2 c N t}$.
On the other side it has a fast decay, $\hat \rho(y) = \frac{y}{2 \sqrt{\pi}} e^{-\frac{y^2}{4}}$
for $y \to +\infty$. 

We have plotted the prediction \eqref{BL} in Fig. \ref{fig:BL} (solid line) where it is also
compared with numerical simulations for various times inside regime II (see below). We see that the agreement of the data 
with the scaling function is excellent. 
\begin{figure}[h!]
\centering
\includegraphics[width=0.7\linewidth]{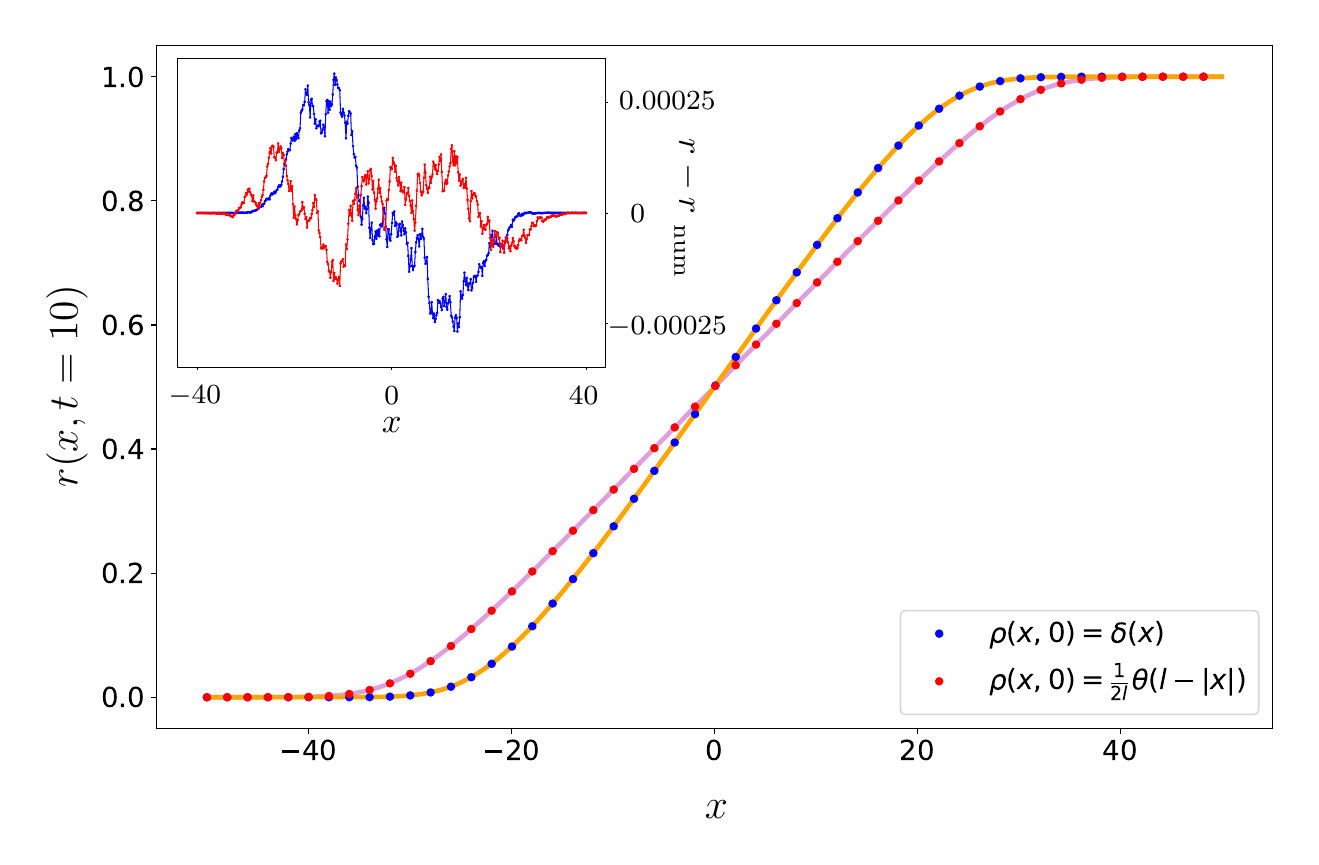}
\caption{The rank field $r(x, t)$ computed numerically (dots) and from the analytical expression (full lines) for two different initial conditions. Red colored dots represent the square initial condition $\rho(x, 0) = \frac{1}{2l}\theta(l-|x|)$ for $l=10$. The numerical data are compared with the analytical expression in Eq. \eqref{burgersbox} (solid purple line). The blue colored dots shows the numerical data for the initial condition where all the particles start from the origin. The analytical expression from Eq. \eqref{burgersdelta} is represented by orange solid line. Inset: point-wise difference between numerical data, $r_{\rm num} (x, t)$, and the analytical expression $r(x, t)$ for both delta and square initial conditions. The deviations are of order $10^{-3}$.
For both plots the rank field is measured at $t=10$ with $\gamma=2$, $T=1$, and $N=1000$. Here the data are averaged over $10^3$ different realizations, which 
shows a small systematic deviation, due to the finite $N$ effects (see Fig. \ref{fig:fl1}).}
\label{fig:rank1}
\end{figure}
In the opposite limit $t \ll \frac{1}{\gamma^2}$ one can check that formula \eqref{burgers1} converges to the Gaussian profile for independent diffusing particles $\rho(x,t) \simeq \frac{1}{\sqrt{4 \pi T t}} e^{-x^2/(4 T t)}$.

We now compare these predictions with a direct numerical calculation of the trajectories $x_j(t)$ from the Langevin equations \eqref{langevin1} where we set $T=1$ and $c= \gamma/N$. In Fig. \ref{fig:time_densities} we study the case where all the particles start from the origin at $t=0$. We plot the numerically evaluated density $\rho(x,t)$ as a function of $x$ for different times $t$ (and several values of $N$), averaged over $10^4$ realizations of the noise. This is compared with the prediction in Eq. \eqref{burgers1} with an initial delta function density. We see that the agreement is quite good,
% While we see a small discrepancy at $t=10/\gamma^2$, the agreement becomes quite good at larger times. The discrepancy at $t=10/\gamma^2$ may originate from the interplay between interactions (which are already at play)
% and the fact that the noise term in \eqref{eqr} being proportional to $\sqrt{\rho(x,t)}$, 
% may still be large for such a time scale. It is quite interesting that 
hence in this time regime the density is very well described by the deterministic Burgers equation even for moderate values of $N$.
%{Indeed, the initial peak in the density quickly disappears and the noise becomes again negligible at large $N$.}

The matching between the solution derived from the Burgers equation and the numerical simulations can be further investigated by looking at the rank fields. In Fig. \ref{fig:rank1} we present the numerical evaluation of the rank fields for $N=1000$ for both the square and the delta initial conditions for $t=50$ and $\gamma=2$. In the inset, we have plotted the difference between the observed and predicted values of the rank field as a function of $x$, as given in
\eqref{burgersbox} and \eqref{burgers1}. One can see that these differences are quite small. 
% although one sees a small systematic
% deviation
% For the square initial condition with $\ell=10$ the agreement is perfect, while we see some small discrepancy for the delta function initial condition. 

Next we study how these differences between the observed and the predicted values of the rank field behave as a function of $N$ and the time $t$, for both initial conditions. For this purpose, we define the error $\Sigma$ as
\be \label{def_Sigma}
\Sigma(N,t) = \sqrt{\frac{1}{n}\sum_{\alpha=1}^{n} (r(y_\alpha,t)-{r}_{\rm num}(y_\alpha,t))^2} \;,
\ee
where $n$ is the number of points on a grid $\{ y_\alpha \}$ at which we numerically computed the rank field (typically $500 \leq n \leq 2000$), $r(y_\alpha,t)$ is the predicted rank field at those points and ${r}_{\rm num}(y_\alpha)$ is the numerical rank field at $y_\alpha$. 
\begin{figure}[t]
\centering
\includegraphics[width=0.5\linewidth]{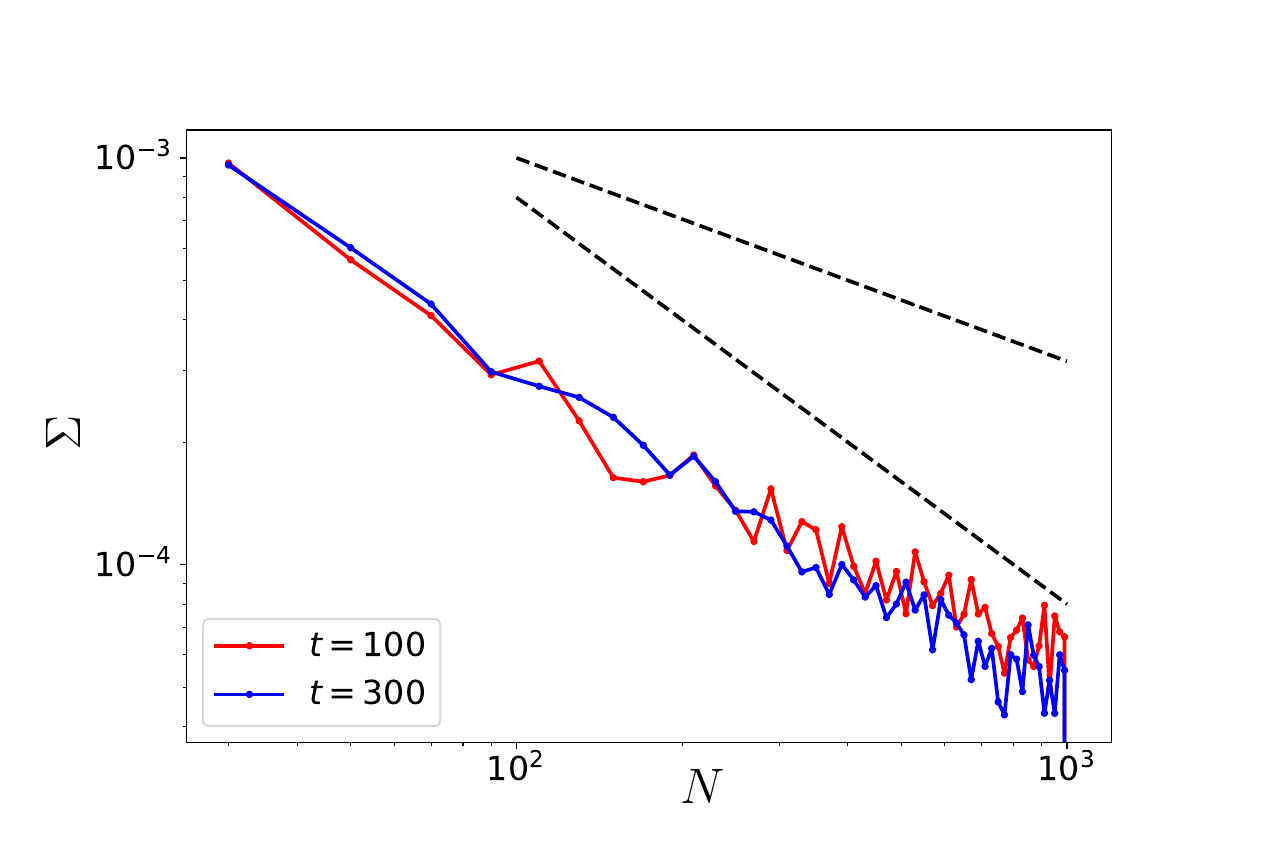}
\caption{Fluctuations of the rank field as measured by $\Sigma(N,t)$ defined in \eqref{def_Sigma}.
We plotted the numerical data for $\Sigma$ as a function of $N$ in log-log scale for $n=500$
and two different times $t=100$ and $t=300$. We have checked that 
the number of points $n$ on the grid [see Eq. (\ref{def_Sigma})], does not affect significantly the data.
Here we fixed $\gamma=1$, $T=1$, and we averaged over $10^4$ realizations of the noise. 
At time zero we start from the square initial condition with $l=10$. 
The dashed lines are guides to the eyes and have slopes respectively $-1/2$ and $-1$. 
% Solid line: fit which shows that the asymptotic is compatible with $\Sigma \sim 1/\sqrt{N}$. 
% We fixed $\gamma=1$, $t\in\{100, 300\}$ and we averaged over $10^4$ realizations of the noise. The grid of points $x_i$ in \eqref{def_Sigma} is chosen
%  inside the interval $[-1000, 1000]$. At time zero we start from square initial condition with $l=10$. {\it Right panel}: The comparison of the fluctuations of the numerical rank field as a function of time for step and delta initial conditions. We fixed $N=500$,$\gamma=1$, $T=1$ and averaged over $10^4$ realizations.
%  The number of grid points is $n=800$. 
} \label{fig:fl1}
\end{figure}
In the left panel of Fig. \ref{fig:fl1}, 
we have plotted $\Sigma(N,t)$, on a log-log scale, as a function of $N$ for different initial profiles at different times. For the square initial condition with $\ell = 10$, we see that $\Sigma(N,t)$ decays with $N$ as a power law $N^{-a}$, where we measure $1/2 < a <1$.
\bigskip

}

%We further confirm this by plotting the fluctuations as a function of time at given $N=500$ in the right panel of  Fig. \ref{fig:fl1}. 

%\begin{figure}[t]
%    \centering
%    \includegraphics[width=0.4\textwidth]{fermisurf.eps}
%    \caption{Representation in phase space $(x,p)$ of the various regimes for the Wigner function for the $1d$ hard box in the limit of a large number of fermions. The thick black rectangle is the Fermi surf (with $k_F = N \pi/2$). The inside of this
%    rectangle is the bulk region (I) where the Wigner function is approximately constant and non zero. Outside of this region it is approximately zero (and strictly zero for $|x|>1$). 
%    The regions where the various crossovers studied in the text take place, i.e., near the Fermi momentum (II), near the hard wall (III), and near the corner (IV) are indicated. The region III extends over a momentum scale of order $O(k_F) = O(N)$, while the region II extends only to order $O(1)$ in momentum space. The blue dashed lines represent schematically the lines of constant value of the Wigner function. }
%    \label{fig:fermisurf}
%\end{figure}
%\vspace*{0.5cm}

%\bibliography{reference}{}
%\bibliographystyle{plain}

\section{Conclusion}

In this paper we have studied the out of equilibrium Langevin dynamics of $N$ particles in one dimension, which interact
only via the linear $1d$ repulsive Coulomb potential, and which are thus allowed to cross. We have focused 
on an initial condition where all the particles are at the origin at time $t=0$. As time increases the gas expands.
We have shown that there are three distinct regimes in time, separated by two characterictic times. In regime I, i.e., 
$t <t_1^*=  T/(c N)^2$ the particles perform essentially independent Brownian diffusion. At $t=t_1^*$,  
the particles start feeling the long-range interaction and for $t> t_1^*$ the size of the gas increases linearly 
$\sim 2 c N t$, this is the beginning of regime II. A plateau forms in the particle density, with boundary layers of size $\sqrt{2 T t}$
whose shape we have explicitly computed using a relation with the Burgers equation. In this regime II the gas is still dense and the
particles still experience many mutual crossings. 
Finally as $t \geq t_2^* = T/c^2$ one enters the regime III where the system is a dilute expanding crystal where the 
particles are well separated. We have studied the regime III at large time thanks to an exact formula obtained from the Bethe
ansatz which we analyzed using a saddle point method. The time dependent particle distribution shows a remarkable analogy with 
the one that describes the equilibrium jellium model in the presence of a quadratic well with a time dependent curvature. 
This allows to quantify the fluctuations of the displacements in the expanding crystal, which are Gaussian and of order $O(\sqrt{t})$ to leading order. Interestingly there are additional $O(1)$ subleading non-Gaussian fluctuations which we obtain exactly. These additional
correlations are purely dynamical in origin and do not have any counterpart in the equilibrium jellium model. 

There are many interesting questions which remain to be studied. One is the role of the initial conditions. Presumably, in the regime III, the 
analogy with the equilibrium jellium model and the leading Gaussian fluctuations are a robust features. 
However it is %quite possible 
{likely that the subleading $O(1)$ non-Gaussian fluctuations for $t \gg T/c^2$ depend on some details
of the initial condition. Indeed we have shown that it is the case for $N=2$, see Appendix \ref{app:init}
where the few lowest cumulants are obtained explicitly for any even initial condition. This shows that the system keeps some memory
of the initial condition even at infinite time. It remains to be investigated how this feature extends to any $N$,
and whether it persists at large $N$}. 
 Another interesting open question is to describe the crossover
from regime II to regime III, i.e., times of order $t_2^* = T/c^2$ or smaller. 
Indeed, as we have shown, the crossover from regime I to regime II for times of order $T/(c N)^2$ can be
described using an hydrodynamic approach based on the Burgers equation. This approach however fails as time increases 
around $t=t_2^*$ when one cannot neglect anymore the discreteness of the particles. In particular
we expect that the boundary layer at the edges of the plateau in the density becomes quite different
from the one computed here. Describing the system in that regime remains an open challenge. 

The stochastic dynamics of long-range interacting systems with a non-equilibrium stationary state, such as the Hamiltonian mean-field model, has been studied in the past \cite{Gupta1,Gupta2}. In contrast, our work concerns the dynamics in a long-range interacting system 
where there is no stationary state. In the present study we have obtained new results in the broader context of interacting Brownian particles
with long range interactions by a combination of analytical methods. A general model much studied recently, in mathematics
and physics, mostly at equilibrium \cite{Lewin,Agarwal_riesz,Beenakker_riesz}, or its quantum generalization \cite{Huse_riesz}
is the so-called Riesz gas in one dimension where the repulsive interaction potential behaves as 
a power law of the distance
$\sim |x_i-x_j|^{-s}$. The case $s=-1$ corresponds to the Coulomb interaction studied here, and the case $s=0$ is the log-gas. 
The out of equilibrium dynamics for general $s$ has not been much addressed (apart of course from $s=0$ and the Dyson's Brownian motion \cite{mehta_book})
with the exception of a very recent work \cite{Mallick_riesz}, where the case $s>0$ was studied using hydrodynamics methods. The present work opens the way for further investigations of non-equilibrium dynamics for Brownian particle systems with 
long range interactions.

    \section*{Acknowledgments}
    PLD and GS thank LPTMS for hospitality. We thank the Erwin Schrödinger Institute (ESI) of the University of Vienna for the hospitality during the workshop {\it Large deviations, extremes and anomalous transport in non-equilibrium systems} in October 2022.
   % This work was supported by ANR grant ANR-17-CE30- 0027-01 RaMaTraF.

\begin{appendix}

\section{More details for two particles $N=2$ }
\label{app:twoparticles}

\subsection{Solution from Laplace transform} 

For two particles one can solve directly the problem in Laplace (as in Supp. Mat. of \cite{PLDRankedDiffusion} but for $c>0$). Noting
$x(t)=\frac{1}{2} (x_1(t) + x_2(t))$ and $y(t)=x_2(t)-x_1(t)$, the center of mass
performs an independent unit Brownian motion, $\dot x(t) = \xi(t)$, while the relative coordinate evolves as
\be 
\dot y = 2 c \, {\rm sgn}(y) + 2 \eta(t) \;,
\ee 
where $\eta(t)$ is an independent unit white noise. 
Its probability density $P(y,t)$ then evolves according to
\be  \label{PP} 
\partial_t P = 2 \partial_y^2 P - 2 c \partial_y ({\rm sgn}(y)  P) \;.
\ee 
starting with the initial condition $P(y,0)=\delta(y)$.
Introducing the Laplace transform $\tilde P(y,s)= \int_0^{+\infty} dt e^{-s t} P(y,t)$ one obtains
\be  \label{equas} 
s \tilde P - P(y,0) = 2 \partial_y^2 \tilde P - 2 c \partial_y ({\rm sgn}(y)  \tilde P) \;,
\ee 
where $P(y,0)$ is the initial condition. It is then easy to solve separately for $y>0$ and $y<0$. There are two integration constants on both sides.
One on each side is set to zero by requiring that $\tilde P(y,s) \to 0$ at $y \to \pm \infty$ for $s>0$. 
Continuity of $P$ at $y=0$ gives another condition and finally, from integrating \eqref{PP} on a small
interval around $y=0$ one obtains the matching condition 
\be  \label{matching} 
P'(0^+,t) - P'(0^-,t) - 2 c P(0,t) = - \frac{1}{2} 
\ee 
and the same relation holds for the Laplace transforms. This leads to the unique solution 
\be \label{PLaplace} 
\tilde P(y,s) = \frac{e^{\frac{c}{2} |y|} }{2(c + \sqrt{c^2 + 2 s})} e^{- \frac{1}{2} \sqrt{c^2+ 2 s} |y|} \;.
\ee 
Hence
\be 
P(y,t)= e^{- \frac{c^2}{2} t} e^{\frac{c}{2} |y|} {\rm LT}^{-1}_{s \to t} 
\frac{1}{2(c + \sqrt{2 s})} e^{- \frac{1}{2} \sqrt{2 s} |y|} \;.
\ee 
Note that the prefactor $e^{- \frac{c^2}{2} t} e^{\frac{c}{2} |y|}$ is exactly the one which appears
in the LL method setting $N=2$. Inverting explicitly the Laplace transform, one finds the formula \eqref{result2} given in the text.
% \be 
% P(y,t)= \frac{e^{-\frac{(|y|-2 c t)^2}{8 t}}}{2 \sqrt{2 \pi }
%   \sqrt{t}}-\frac{1}{4} c e^{c |y|}
%   \text{erfc}\left(\frac{2 c t+|y|}{2 \sqrt{2}
%   \sqrt{t}}\right)
% \ee 

\subsection{Comparison with the general $N$ formula}

Let us rewrite the general formula \eqref{exactN} for $N=2$, denoting
$y=x_2-x_1$ and $x=\frac{x_1+x_2}{2}$, i.e., $x_2=x+y/2$ and $x_1=x-y/2$ as above
(with $dx_1dx_2=dx dy$). 
It reads for $x_1 \leq x_2$, that is for $y \geq 0$
\be \label{exactN2} 
P(x_1,x_2,t) = e^{\frac{c}{2} |y| }  e^{- \frac{c^2}{2} t} 
\int_{\mathbb{R}} \frac{dk_1}{2 \pi} \int_{\mathbb{R}}  
\frac{dk_2}{2 \pi} \frac{i k_1- i k_2}{i k_1 - i k_2 + c} \, 
e^{- t (k_1^2 + k_2^2) + i x (k_1+k_2) + \frac{i}{2} y (k_2-k_1)} 
\ee 
Let us denote $k_1+k_2=k$ and $q = (k_2-k_1)/2$. The above expression factorizes and one obtains %for $y \leq 0$ 
\be \label{exactN22} 
P(x_1,x_2,t) = e^{\frac{c}{2} |y| }  e^{- \frac{c^2}{2} t} 
\left( \int_{\mathbb{R}} \frac{dk}{2 \pi} 
e^{- t \frac{k^2}{2}   + i x k }  \right) \times  \left( 
 \int_{\mathbb{R}}  \frac{dq}{2 \pi} \frac{- i 2 q}{- i 2 q + c}  \, 
e^{- 2 t q^2  + i y q } \right) 
\ee 
Hence the center of mass motion decouples and one has
\be 
P(x_1,x_2,t) dx_1 dx_2 =  \frac{1}{\sqrt{2 \pi t}} e^{-\frac{x^2}{2 t}} P(y,t) dx dy
\ee 
with for $y \geq 0$
\be \label{py} 
P(y,t)=  e^{\frac{c}{2} |y| }  e^{- \frac{c^2}{2} t} \int_{\mathbb{R}}  \frac{dq}{2 \pi} \frac{- i 2 q}{- i 2 q + c}  \, 
e^{- 2 t q^2  + i y q }
\ee 
The Laplace transform of this expression w.r.t. time $t$ reads
\be 
\tilde P(y,s) =  e^{\frac{c}{2} |y| }   \int_{\mathbb{R}}  \frac{dq}{2 \pi} \frac{- i 2 q}{- i 2 q + c}  \,  
\frac{1}{s + 2 q^2 + \frac{c^2}{2}} 
e^{i y q }
\ee 
One has $s + 2 q^2 + \frac{c^2}{2} = 2 (q-q_+)(q-q_-)$ with $q_\pm = \pm \frac{i}{2} \sqrt{2 s + c^2}$.
Since here $y>0$ we must close the contour in the upper half-plane. The pole at $q=- i c/2$ thus 
does not contribute, and the pole at $q=q_+$ contributes. We obtain from its residue
\be 
\tilde P(y,s) =  i e^{\frac{c}{2} |y| }  
\frac{1}{2 (c + \sqrt{2 s + c^2} )} \frac{- 2 i q_+}{(q_+-q_-)} e^{- \frac{y}{2} \sqrt{2 s + c^2}} 
= e^{\frac{c}{2} |y| }  
\frac{1}{2 (c + \sqrt{2 s + c^2} )} e^{- \frac{y}{2} \sqrt{2 s + c^2}} 
\ee 
which, using that the final result must be even in $|y|$ recovers the previous result, see \eqref{Laplace2} and \eqref{PLaplace}. 

\subsection{Cumulants of $|y|- 2 c t$} \label{sec:app_cumul}

To obtain the cumulants of $|y|- 2 c t$, one can compute the Laplace transform of
the generating function
\be 
\int_0^{+\infty} dt \int dy P(y,t) e^{\lambda (|y|- 2 c t) - s t } = \int dy \tilde P(y,s+ 2 c \lambda) e^{\lambda |y|} 
\ee 
Using the expression \eqref{Laplace2}, expanding in $\lambda$ and performing the Laplace inversion,
we find the moments, and taking the logarithm we obtain the cumulants. From now on we set $c=1$ for simplicity and will restore it in the text.
One obtains
\bea 
&& \langle |y| \rangle - 2 t = (t+1)
   \text{erf}\left(\frac{\sqrt{t}}{\sqrt{2}}\right)-t+
   \sqrt{\frac{2}{\pi }} e^{-t/2} \sqrt{t} = 1 + O(t^{-3/2} e^{-t/2}) \\
&& \langle (|y|-2  t)^2 \rangle = 2 \left(t^2+1\right)
   \text{erfc}\left(\frac{\sqrt{t}}{\sqrt{2}}\right)-2
   \sqrt{\frac{2}{\pi }} e^{-t/2} \sqrt{t} (t-1)+4 t-2 = 4 t - 2 + O(t^{-1/2} e^{-t/2}) \\
&& 
\langle (|y|-2  t)^3 \rangle  = 12 t + 6  + O(t^{1/2} e^{-t/2}) \\
&& 
\langle (|y|-2  t)^4 \rangle  = 48 t^2 - 48 t - 24  + O(t^{3/2} e^{-t/2}) \\
&&\langle (|y|-2  t)^5 \rangle  = 240\,t^2 + 240\, t + 120 + o(1) \\
&&\langle (|y|-2  t)^6 \rangle = 960\,t^3 - 1440 t^2 - 1440 t - 720 + o(1) \;,
\eea
The first four cumulants are given in the text, and we further obtain by this method the next two cumulants 
% %
% {\blue
% Higher orders moments are given, for large $t$, by
% \bea \label{higher_mom}
% &&\langle (|y|-2  t)^5 \rangle  \simeq 240\,t^2 + 240\, t + 120 \\
% &&\langle (|y|-2  t)^6 \rangle \simeq 960\,t^3 - 1440 t^2 - 1440 t - 720 \;,
% \eea
% from which we get the higher order cumulants
\be \langle (|y|-2 c t)^5 \rangle_c  \simeq 744 + o (1) \quad , \quad 
\langle (|y|-2 c t)^6 \rangle_c \simeq - 7560 + o (1) \;. \label{higher_cum}
\ee
where here and above the terms $o(1)$ are exponentially small corrections in $t$. We see that although the moments are polynomial in
$t$ the cumulants are simply $O(1)$. Furthermore, there are no power law corrections of the type $1/t^p$, $p \geq 1$,
to either moments or cumulants. Finally, one can check that the cumulants obtained exactly here 
up to order $6$ agree with the general prediction obtained in Section \ref{subsec:cumulants} 
for $k\geq 3$ (restoring $c$) 
\bea \label{cumul_gen}
 \langle (|y|-2 c t)^k \rangle_c = \frac{1}{c^k}\, (-1)^{k-1} (k-1)! (2^k-1) + o (1) \;.
\eea

\subsection{More general initial conditions}
\label{app:init}

In the Supp. Mat. of \cite{PLDRankedDiffusion} the solution for $P(y,t)$ for $N=2$ was obtained for
a more general class of initial conditions. The calculation was performed for $c<0$ (i.e., setting $c=-1$), here
we adapt it for $c>0$, following the same steps. 

One considers an initial condition $P(y,0)$ which is smooth around $y=0$ (for convenience) and
an even function of $y$. Then $P(y,t)$ is also smooth around $y=0$ and an even function of $y$.
One again defines the Laplace transform with respect to time (w.r.t.),
$\tilde P(y,s)= \int_0^{+\infty} dt e^{-s t} P(y,t)$, and one also defines
the half-sided Laplace transform w.r.t. space,
$\hat P(\mu,s)= \int_0^{+\infty} dy \, e^{- \mu y} \tilde P(y,s)$ (thus using $\lambda=-\mu$ as compared
to the notations used in the previous section).
One also denotes $P_0(\mu)= \int_0^{+\infty} dy \, e^{- \mu y} P(y,0)$, the half-sided Laplace transform of the initial
condition. The normalization condition on the half-space implies that 
$P_0(0)=1/2$ and {$\hat P(0,s)=\frac{1}{2 s}$}.
Taking the double Laplace transform of Eq. \eqref{equas} w.r.t. $x$ and $t$ then leads to
\be \label{double_LT}
s \hat P(\mu,s) - P_0(\mu) = 2 \mu (\mu-c) \hat P(\mu,s) - 2 \tilde P'(0,s) - 2 (\mu-c) P(0,s) \;.
\ee 
Integrating (\ref{double_LT}) around $y=0$ leads to the jump conditions $P'(0^+,s)- c P(0^+,s)=0$ and its solution reads
\be 
\hat P(\mu,s)= \frac{P_0(\mu) - 2 \mu \tilde P(0,s)}{s - 2 \mu(\mu-c)} \;.
\ee 
We will use the same condition as was used for $c=-1$ in \cite{PLDRankedDiffusion} to determine the unknown function $\tilde P(0,s)$, namely that the residue of the pole at $s=2 \mu(\mu-c)$ should vanish. Indeed, for $\mu>c$, $2 \mu(\mu-c) >0$ and a pole at $s>0$ would lead 
to a growing exponential in time, which is excluded. Hence one has
\be 
\tilde P(0,s= 2 \mu(\mu-c)) = \frac{P_0(\mu)}{2 \mu} \;.
\ee 
Equivalently, setting $\mu= \mu_s= \frac{1}{2} (c + \sqrt{c^2+2 s})$ (the positive root) one must have
\be 
\tilde P(0,s) = \frac{P_0(\mu_s=\frac{1}{2} (c+\sqrt{c^2+2 s}))}{c+ \sqrt{c^2+2 s}} 
\ee 
The solution is thus
\be 
\hat P(\mu,s)= \frac{1}{s - 2 \mu(\mu-c)} 
 \left( P_0(\mu) - 2 \mu \frac{P_0(\frac{1}{2} (c+\sqrt{c^2+2 s}))}{1+ \sqrt{c^2+2 s}} \right) \;.
\ee 

Let us now compute the cumulants of $|y|- 2 c t$. We again use the 
Laplace transform in time of the cumulant generating function
\be 
\int_0^{+\infty} dt \int_{-\infty}^{+\infty} dy P(y,t) e^{-\mu (|y|- 2 c t) - s t } = \int_{-\infty}^{+\infty} dy \, \tilde P(y,s- 2 c \mu) e^{- \mu |y|} 
= 2 \hat P(\mu,s - 2 c \mu)  \;.
\ee 
Let us set from now on $c=1$ (which means lengths are in units of $1/c$ and time of $1/c^2$).
Let us denote $m_k(t)=\langle (|y|- 2 c t)^k \rangle$ the moments
and $\tilde m_k(s)$ their Laplace transform in time, and $\kappa_k(t)$ the cumulants. One has
\be 
\tilde m_k(s) = 2 (-1)^k \partial_\mu^k|_{\mu=0} \hat P(\mu,s - 2  \mu)  \;.
\ee 
To extract the large time behavior we perform the small $s$ expansion for each moment.
For instance one finds
\be 
\tilde m_1(s)= \frac{2}{s} (\hat P_0(1) - \hat P_0'(0)) + \hat P_0'(1) - \hat P_0(1) + O(s) \;,
\ee 
which implies that 
\be 
m_1(t) = 2 (\hat P_0(1) - \hat P_0'(0))  + f_1(t) \quad , \quad \int_0^{+\infty} dt \, f_1(t) = \hat P_0'(1) - \hat P_0(1) \;,
\ee 
where $f_1(t)$ decays to zero at infinity. Hence the $O(1)$ constant in the first cumulant reads
\be 
\kappa_1(t = +\infty) = 2 (\hat P_0(1) - \hat P_0'(0))  = \langle y + e^{-y}  \rangle_0 
\ee 
where $\langle \dots \rangle_0$ means the average with respect to the initial condition $P(y,0)$. 
One recovers $\kappa_1(t = + \infty) =1$ in the limit where $P(y,0)=\delta(y)$. We see that 
$\kappa_1(t= +\infty)$ depend on the initial condition.
Next one obtains
\be 
\tilde m_2(s)= \frac{4}{s^2} + \frac{2}{s} \left( 2   \hat P_0'(1)+ \hat  P_0''(0)-2   \hat P_0(1) \right) +O(s^0 )
\ee 
which leads to
\be 
m_2(t) 
= 4 t + \langle y^2 - 2 (1+y) e^{-y} \rangle_0 + o(t) \;.
\ee 
Thus we find that the $O(1)$ constant in the second cumulant $\kappa_2(t) = 4 t + \tilde \kappa_2(t)$ reads
\be 
\tilde \kappa_2(t = + \infty) =   \langle y^2 - 2 (1+y) e^{-y} \rangle_0 - \langle y + e^{-y}  \rangle_0^2 
\ee 
and one recovers $\tilde \kappa_2(t=+\infty) =-3$ in the limit where $P(y,0)=\delta(y)$. 
The third and fourth moments are
\bea 
&& m_3(t) = 12 t \, \langle y+ e^{-y} \rangle_0 + \langle y^3+3 e^{-y} \left( y^2+2 y+2\right) \rangle_0  + o(t) \\
&& m_4(t) =
48 t^2+ 24 t \langle y^2- 2 e^{-y}   (y+1) \rangle + \langle y^4-4 e^{-y}  \left(y^3+3 y^2+6 y+6\right) \rangle + o(t) 
\eea 
from which one obtains the third and fourth cumulants, which have heavy expressions
not displayed here. One checks that all positive orders in $t$
cancel in the $k$-th cumulant, $k \geq 3$, which thus 
goes to a $O(1)$ constant, $\kappa_k(t=+\infty)$, as large time. These
$O(1)$ constants carry information, up to infinite time, 
about some details of the initial condition.

{
\section{More on the cumulants from the saddle point} 

The saddle point method used in Section \ref{subsec:cumulants} would predict power law in time
corrections to the cumulants, but already for $N=2$ we know that these do not exist. Let us
focus on $N=2$. 
To understand this apparent paradox, let us go one step back and start again from the formula \eqref{py} (setting $c=1$ for simplicity here)
\be \label{py2} 
\langle e^{\lambda y} \rangle = 2 \int_{0}^{+\infty} dy e^{(\lambda + \frac{1}{2})  y }  e^{- \frac{1}{2} t} 
\int_{\mathbb{R}}  \frac{dq}{2 \pi} \frac{- i 2 q}{- i 2 q + 1}  \, 
e^{- 2 t q^2  + i y q }
\ee 
Instead of performing the saddle point on $q$ and then perform the saddle point on the resulting expression (as we did in Section \ref{subsec:cumulants}), let
us simply rewrite \eqref{py} using the shifted variables $y = 2 t (1+ 2 \lambda) + \hat y$ and $q=\frac{i}{2} (1+ 2 \lambda)+ \hat q$.
One obtains
\be \label{py3} 
\langle e^{\lambda y} \rangle = 2 \int_{- {2}t (1+2 \lambda)}^{+\infty} d\hat y 
\int_{\mathbb{R}}  \frac{d\hat q}{2 \pi} \frac{1+ 2 \lambda - 2 i \hat q}{2+ 2 \lambda - 2 i \hat q}
e^{2 \lambda t + 2 \lambda^2 t - 2 t \hat q^2  + i \hat y \hat q}  
\ee 
Note that the integration contour of $\hat q$ was $\mathbb{R}- \frac{i}{2} (1+ 2 \lambda)$ but we
brought it back to $\mathbb{R}$ since the pole at $q=- i/2$ is not crossed along the way (provided
$1+2 \lambda>0$). Note that this formula is {\it exact}, no saddle point has been made. Now we split the integral over $\hat y$ in two pieces,
i.e., we write $\int_{- 2 t (1+2 \lambda)}^{+\infty} d\hat y = \int_{- \infty}^{+\infty} d\hat y  - \int_{-\infty}^{- 2 t (1+2 \lambda)} d\hat y$.
In the first piece we use $\int_{- \infty}^{+\infty} d\hat y e^{i \hat y \hat q} = 2\pi \delta(q) $ and we obtain
(formally the second piece corresponds to $y<0$ in \eqref{py2})
\be \label{py4} 
\langle e^{\lambda y} \rangle = e^{2 \lambda t + 2 \lambda^2 t}  \left( 
 \frac{1+ 2 \lambda}{1+ \lambda}  
 - 2 \int_{-\infty}^{- 2 t (1+2 \lambda)} d\hat y \int_{\mathbb{R}}  \frac{d\hat q}{2 \pi} \frac{1+ 2 \lambda - 2 i \hat q}{2+ 2 \lambda - 2 i \hat q}
e^{ - 2 t \hat q^2  + i \hat y \hat q} \right)
\ee 
The idea is that the second piece is exponentially small at large time. For instance for $-2 t (2 + 2 \lambda) < \hat y < - 2 t (1+2 \lambda) $
one can evaluate the integral over $\hat q$ by a saddle point
method, with a saddle point at $\hat q = i \frac{\hat y}{4 t}$. This leads to 
\be \label{py5} 
\int_{\mathbb{R}}  \frac{d\hat q}{2 \pi} \frac{1+ 2 \lambda - 2 i \hat q}{2+ 2 \lambda - 2 i \hat q}
e^{ - 2 t \hat q^2  + i \hat y \hat q} \simeq \frac{1}{\sqrt{8 \pi t}}  e^{- \frac{\hat y^2}{8 t} }  
 \frac{1+ 2 \lambda + \frac{\hat y}{2 t} }{2+ 2 \lambda  + \frac{\hat y}{2 t} }
\ee 
A similar estimate can be obtained for $\hat{y}< -2 t (2 + 2 \lambda)$. Hence the final integral over $\hat y$
is dominated by its upper bound and is thus of order $e^{- \frac{t}{2} (1+ 2 \lambda)^2}$ with algebraic prefactors. This
gives an exponentially small correction to the cumulants, of order $e^{- \frac{c^2 t}{2}}$ (restoring $c$), which
is indeed what is obtained by an exact calculation. Note that the exponentially small correction term in \eqref{py4}
comes from trajectories which cross each other, which become subdominant for $c^2 t \gg 1$. 
Finally, these calculations can be generalized to any $N$, although we will display it here. 

}

\section{Asymptotic behavior of the solution of Burger's equation} \label{app:bl}

To study the boundary layer of the density $\rho(x,t)$ in (\ref{burgers1}) near its right edge, let us recall some useful formulae for the delta initial condition, namely
\be \label{burgersdelta2} 
 r(x,t)= - \frac{T}{\gamma} \partial_x \log \left( f(x,t) + f(-x,t) \right) \quad , \quad  f(x,t) = e^{- \frac{\gamma x}{2 T} } 
   \text{erfc}\left(\frac{\gamma t-x}{2
   \sqrt{T t}}\right) \;, 
\ee
which we have slightly simplified. Let us focus near the right edge at $x = \gamma t$, 
and set $x = \gamma t + y \sqrt{T t}$. Then we find
\bea     
&& f(x,t) = e^{- \frac{\gamma^2 t}{2 T} - \frac{\gamma y \sqrt{t}}{2 \sqrt{T}} } {\rm erfc}(- \frac{y}{2}) \\
&& f(-x,t) = e^{- \frac{\gamma^2 t}{2 T} - \frac{\gamma y \sqrt{t}}{2 \sqrt{T}} } e^{-\frac{y^2}{4}} 
\frac{\sqrt{T}}{\gamma \sqrt{\pi t} } \left(1 - \frac{y \sqrt{T}}{2 \gamma \sqrt{t}}+ O(\frac{1}{t}) \right) \;.
\eea   
Hence for $t \gg T/\gamma^2=T/(c N)^2$ and $y=O(1)$ we see that the first term dominates. Hence in that limit
one has
\bea   
r(x,t)= \frac{1}{2} - \frac{\sqrt{T}}{\gamma \sqrt{t}} \partial_y \log \left( {\rm erfc}(- \frac{y}{2} ) \right) 
= \frac{1}{2} -\frac{\sqrt{T}}{\gamma \sqrt{t}} \hat r(y) \quad , \quad \hat r(y) =  \frac{e^{-\frac{y^2}{4}}}{\sqrt{\pi}   \text{erfc}\left(-\frac{y}{2}\right)}
\eea    
which describes the boundary layer at the right edge. It behaves as $\hat r(y) \simeq \frac{y}{2} + \frac{1}{y}$ for $y \to - \infty$, hence matches the
linear behavior of the plateau. The density at the edge thus takes the following boundary layer form
\be   \label{BL_app}
\rho(x,t) = \frac{1}{\gamma t} \hat \rho(\frac{x-\gamma t}{\sqrt{T t}}) 
\quad , \quad \hat \rho(y) = \frac{e^{-\frac{y^2}{2}}  \left(2+\sqrt{\pi } e^{ \frac{y^2}{4}} y \, \text{erfc}\left(-\frac{y}{2}\right)\right)}{2 \pi \text{erfc}\left(-\frac{y}{2}\right)^2}
\ee   
where the scaling function has the asymptotic behaviors
\bea     
&& \hat \rho(y) = \frac{1}{2} - \frac{1}{y^2} + O(\frac{1}{y^4}) \quad , \quad y \to -\infty \\
&& \hat \rho(y) = \frac{y}{2 \sqrt{\pi}} e^{-\frac{y^2}{4}} \quad , \quad y \to +\infty
\eea    
The boundary layer form of the density thus matches the density of the plateau $\rho(x,t) \simeq \frac{1}{2 \gamma t}= \frac{1}{2 c N t}$.

\end{appendix}

\end{document}